\documentclass[twocolumn,journal]{IEEEtran}

\IEEEoverridecommandlockouts

\usepackage{color}
\usepackage{graphicx}
\usepackage{epstopdf}
\usepackage{amsmath}
\usepackage{amssymb}
\usepackage{algorithm}
\usepackage{algorithmic}
\usepackage{amsmath}
\usepackage{multirow}
\usepackage{booktabs}
\usepackage{array}
\usepackage{amsthm}
\usepackage{stfloats}
\usepackage{caption}
\usepackage{subfigure}
\usepackage{bm}
\usepackage{setspace}
\usepackage{mdframed}
\usepackage{diagbox}
\usepackage{balance}

\newcommand{\be}{\begin{equation}}
\newcommand{\ee}{\end{equation}}
\newcommand{\bea}{\begin{eqnarray}}
\newcommand{\eea}{\end{eqnarray}}
\newcommand{\ba}{\begin{array}}
\newcommand{\ea}{\end{array}}

\title{A Hierarchical DRL Approach for Resource Optimization in Multi-RIS Multi-Operator Networks
\thanks{Manuscript received 13 October 2024; revised 2 January 2025; accepted 19 February 2025. This work is supported in part by the National Natural Science Foundation of China (Grant No. 62371090, 62471086, and 62271163), in part by Fundamental Research Funds for the Central Universities (Grant No. DUT24ZD125 and DUT24RC(3)005), and in part by Liaoning Applied Basic Research Program (2023JH2/101700364). The associate editor coordinating the review of this article and approving it for publication was Dr. Xiaowen. Gong. \emph{(Corresponding author: Ming Li; Wei Wang.)}}
\thanks{H. Zhang, M. Li, and W. Wang are with the School of Information and Communication Engineering, Dalian University of Technology, Dalian 116024, China (e-mail: dlutzhc@mail.dlut.edu.cn; mli@dlut.edu.cn; wangwei2023@dlut.edu.cn).}
\thanks{H. Zhou was with the Department of Computer Science, McGill University, Montr\'eal, QC, H3A 0G4, Canada. e-mail: hzhou098@uottawa.ca}
\thanks{Z. Lu is with Beijing University of Posts and Telecommunications, Beijing 100876, China, and State Key Laboratory of Wireless Mobile Communications (CICT), Beijing 100191, China (e-mail:luzp@cict.com).} 
}
\author{ \vspace{-20pt}
Haocheng Zhang,
    Wei Wang,~\IEEEmembership{Senior Member,~IEEE,}
      Hao Zhou,~\IEEEmembership{Member,~IEEE,}
        Zhiping Lu,\\
       and Ming Li,~\IEEEmembership{Senior Member,~IEEE}
\vspace{-20pt}
}
\begin{document}
\maketitle
\pagestyle{empty}
\thispagestyle{empty}

%\doublespacing

\begin{abstract}
As reconfigurable intelligent surfaces (RIS) emerge as a pivotal technology in the upcoming sixth-generation (6G) networks, its deployment within practical multiple operator (OP) networks presents significant challenges, including the coordination of RIS configurations among OPs, interference management, and privacy maintenance. A promising strategy is to treat RIS as a public resource managed by an RIS provider (RP), which can enhance resource allocation efficiency by allowing dynamic access for multiple OPs.  However, the intricate nature of coordinating management and optimizing RIS configurations significantly complicates the implementation process. In this paper, we propose a hierarchical deep reinforcement learning (HDRL) approach that decomposes the complicated RIS resource optimization problem into several subtasks. Specifically, a top-level RP-agent is responsible for RIS allocation, while low-level OP-agents control their assigned RISs and handle beamforming, RIS phase-shifts, and user association. By utilizing the semi-Markov decision process (SMDP) theory, we establish a sophisticated interaction mechanism between the RP and OPs, and introduce an advanced hierarchical proximal policy optimization (HPPO) algorithm. Furthermore, we propose an improved sequential-HPPO (S-HPPO) algorithm to address the curse of dimensionality encountered with a single RP-agent. Experimental results validate the stability of the HPPO algorithm across various environmental parameters, demonstrating its superiority over other benchmarks for joint resource optimization. Finally, we conduct a detailed comparative analysis between the proposed S-HPPO and HPPO algorithms, showcasing that the S-HPPO algorithm achieves faster convergence and improved performance in large-scale RIS allocation scenarios.
\end{abstract}


%\newpage
%\maketitle
%\vspace{-0.3 cm}
\section{Introduction}\label{section:1}

Reconfigurable intelligent surface (RIS) is regarded as a key technology for improving communication quality in the upcoming sixth-generation (6G) networks \cite{CM Pan}. 
By carefully designing the phase-shifts of passive elements, RIS can dynamically adjust the transmission characteristics of incident electromagnetic waves, thereby reshaping the radio propagation environment \cite{TWC Wu}.  When the line-of-sight (LoS) signal strength between the base station (BS) and the user is insufficient,  deploying RISs can enable a coherent combination of multi-path signals, thereby enhancing the received signal power and improving the user’s quality of service (QoS). Additionally, RIS offers several attractive advantages, such as low energy consumption, reduced hardware costs, and ease of installation, which have encouraged researches into its integration with other key technologies \cite{JSTSP Liu,TWC He}.

Efficient RIS resource allocation plays a crucial role in taking full advantage of RIS-aided wireless networks, e.g., selectively enhancing specific BS-user links to improve system spectral/energy efficiency \cite{TVT Sun}-\cite{TVT Liu}. Specifically, the authors in \cite{TVT Sun} computed the signal-to-leakage-plus-noise ratio for all possible RIS-user associations and applied a graph-based algorithm to solve the bilateral association problem. In \cite{TVT Luong}, the authors developed an evolutionary game strategy based on utility functions to enable users to adaptively select RISs. In \cite{TVT Liu}, the authors proposed a combined majorization minimization and alternating direction method of multipliers algorithm to jointly design the BS-RIS-user association, beamforming, and RIS phase-shifts. It is noteworthy that the aforementioned studies mainly focus on single network-operator (OP) scenarios, where global channel state information (CSI) is shared among all BSs to allocate RIS resources.

However, in real-world cellular networks, multiple  OPs often coexist, making it unrealistic to assume that all OPs will share private information, such as user locations, traffic types, and  CSI. Nevertheless, the strategy of each OP independently utilizing RIS resources introduces significant challenges in multi-OP network environments. Consequently, the multi-RIS and multi-OP scenario poses substantial challenges for practical RIS deployments, where each OP independently manages and controls RISs based on local observations.
On the one hand, due to time-varying traffic distribution and interference, the demand for RISs from different OPs fluctuates dynamically \cite{CST Xu}. Thus, the approach that each OP optimizes RISs separately will not only lead to lower resource utilization efficiency, but also substantially increase the deployment and maintenance costs. On the other hand, the independent configuration of RISs may significantly increase the interference among users belonging to different OPs, necessitating designs to balance trade-offs among all OPs \cite{GC Angjo}. To this end, a practical solution is to consider RISs as public resources that are deployed and managed by a third-party entity designated as the RIS provider (RP) \cite{ICC Frisanco}.

Some recent studies have investigated the coupling effects of RISs operated by different OPs, examining how interactions among multiple RISs can influence overall system performance. In \cite{TCOM Gurgunoglu}, the authors explored the impact of RIS on pilot contamination during the channel estimation process for OPs. In \cite{GC Lin}, the authors designed a block coordinate descent framework for resource optimization, which is based on the assumption of shareable CSI. In \cite{WCL Yashvanth}, the authors analyzed the growth trend of the optimal spectral efficiency of non-associated OP and concluded that RIS always benefits non-associated OP. The authors in \cite{TCOM Yashvanth}-\cite{WCL Miridakis} considered the inter-OP interference caused by RIS phase-shifts. Nevertheless, these studies assume that RIS is always constantly associated with a specific OP and overlook its characteristics in multi-band networks. Given the frequency-selective properties of RIS, it offers tunable phase-shifts only for the transmission signals of its associated OP, while applying nearly fixed phase-shifts to signals in the frequency bands of unassociated OPs \cite{TCOM Cai}-\cite{TCOM Li}. This implies that improper allocation of RIS by the RP can significantly degrade the BS-RIS-user cascaded channel, thereby reducing the efficiency of OPs' resource optimization, such as beamforming design and user association.

Given the above opportunities and challenges, it is crucial to design an efficient algorithm to effectively solve the joint optimization problem involving the RP and OPs. Unfortunately, traditional centralized algorithms are impractical for real-world deployment. On the one hand, conventional iterative optimization exhibits high computational complexity, which cannot meet the demands of real-time resource optimization. On the other hand, the shareable information between the RP and each OP is limited \cite{TN Pan}. In multi-RIS multi-OP networks, public RIS resources and the resources of individual OPs are tightly coupled. However, the operation information required---such as RIS-related CSI, resource allocation schemes, and user QoS of all OPs---cannot be centrally processed in real-time. As a result, the joint RIS resource optimization problem needs to be executed in a distributed manner by both the RP and OPs. Latest studies have focused on deploying non-cooperative auction algorithms for RIS management within the RP,  while overlooking the joint optimization of OP resources \cite{TN Pan}, \cite{CL Schwarz}. Besides, auction algorithms often require substantial additional computational overhead to obtain bidding information \cite{TVT Cai}. Consequently, the limitations of traditional approaches prompt us to explore machine learning (ML) techniques, which establish functional mappings for real-time inference, offering performance advantages in non-convex problems \cite{CST Hussain}-\cite{TWC Zhang}.

In this paper, we introduce a hierarchical deep reinforcement learning (HDRL) approach to handle the challenges in the multi-RIS multi-OP networks. As a branch of ML, deep reinforcement learning (DRL) optimizes model parameters in parallel through real-time interactions, further reducing training costs and data processing overhead \cite{JSAC Huang}-\cite{CST Zhou}. Building on this foundation, the benefits of utilizing HDRL  are primarily evident in three key aspects: 1) Firstly, traditional DRL algorithms rely on global states to train an optimal policy, which can only address the joint optimization problem in a centralized manner. In contrast, HDRL can decompose this problem into distributed tasks, allowing the RP and individual OPs to solve them independently \cite{TCCN Zhou}, \cite{TCOM Geng}. 2) Secondly, guided by semi-Markov decision process (SMDP) theory, HDRL establishes a hierarchical structure between the RP and OPs, where the top-level RP-agent is responsible for RIS allocation, while the low-level OP-agents focus on their own resource optimization. This generalized framework effectively reduces the complex coordination requirements. 3) Finally, by leveraging the feature extraction capability of the neural network, OP-agents can optimize their resources through implicit environmental information, significantly reducing the overhead of RIS-related channel estimations \cite{WCL Guo}-\cite{TCOM Yang}. At the same time, the RP-agent can use implicit information uploaded by OP-agents for RIS allocation, thus avoiding the leakage of operational privacy information.

In addition, the proposed HDRL approach distinguishes itself from most existing works in the field of communications. 1) Firstly, in \cite{TCOM Geng} and \cite{TWC Wang}, HDRL has been applied solely to optimize a single entity for deploying heterogeneous frequency tasks and enhancing the performance of centralized DRL. In contrast, we utilize HDRL to construct a joint resource optimization framework for both the RP and OPs, which imposes higher demands on the design of interaction mechanisms between different entities. 2) Secondly, HDRL is not simply a random combination of model-free DRL algorithms as in \cite{TCCN Zhou}, \cite{TCOM AL-Eryani}. Instead, it is meticulously selected based on real-time environmental conditions and extensive simulation experiments. 3) Finally, we evaluate the impact of the high-dimensional output space at the RP-level on the overall performance of HDRL and propose an improvement scheme, which has not been considered in previous works \cite{TCCN Zhou, TCOM Geng, TWC Wang, TCOM AL-Eryani}. 
To this end, we deploy an improved sequential optimization technique to lower the dimensional complexity.  In summary, the proposed HDRL approach specifically addresses the challenging joint resource optimization problem in multi-RIS-assisted multi-OP networks. The main contributions of this paper can be summarized as follows.

\begin{itemize}

\item We consider a multi-RIS multi-OP network, where each OP consists of multiple BSs and multiple users, and each RIS exhibits frequency-selective characteristics while being coordinated by a central RP. We formulate the joint resource optimization problem for the RP and OPs. Specifically, the RP aims to maximize the sum-rate of all OPs, which is jointly influenced by RIS allocation, RIS phase-shifts, beamforming, and user association.

\item Next, we propose an HDRL framework to decompose the joint problem into several subtasks, allowing the RP and OPs to optimize resources independently. Specifically, the top-level RP first performs RIS allocation, while each lower-level OP controls the assigned RIS and designs internal beamforming, user association, and RIS phase-shifts. To prevent privacy breaches, we carefully define the key elements of the HDRL framework and enhance the interaction mechanisms between the RP and OPs.

\item Furthermore, based on the HDRL framework, we introduce the hierarchical proximal policy gradient (HPPO) algorithm to facilitate real-time policy updates. In HPPO algorithm, RP-agent samples from the output probabilities to generate the RIS allocation, while OP-agents sample from the output Beta function. Besides, to address potential dimensionality issues in large-scale RIS allocation, we further decompose the RP-level MDP  and design an improved sequential-HPPO (S-HPPO) algorithm.

\item  Finally, simulation results indicate that HPPO algorithm effectively facilitates the RP's allocation of RISs, thereby assisting each OP to further enhance the sum-rate of subscribers. Compared to various RIS allocation benchmarks in the RP, the HPPO algorithm achieves the best performance. Additionally, S-HPPO algorithm effectively addresses the dimensionality curse encountered by HPPO in large-scale RIS allocation, exhibiting faster convergence and superior performance.
\end{itemize}

\section{System Model and Problem Formulation} \label{section:2}
\begin{figure*}[t]
\centering
\includegraphics[width = 7 in]{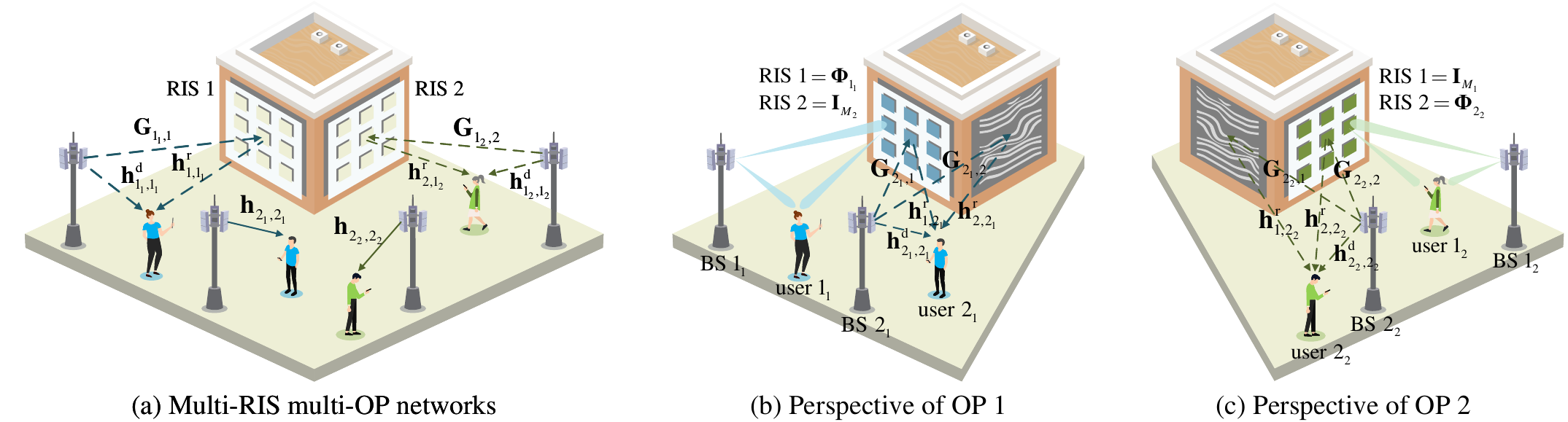}
\caption{A schematic of the multi-RIS assisted multi-OP system with $L$=2, $S$=2, $J_s$=2, and $K_s$=2. RIS 1 and RIS 2 are allocated by RP to OP 1 and OP 2, respectively.}
\label{fig_system}
\vspace{-0.3 cm}
\end{figure*}

As shown in Fig. \ref{fig_system}(a), we consider a downlink communication network with total $S$ OPs and $L$ RISs. The sets of all OPs and RISs are denoted as $\mathcal{S}=\{1,2,...,S\}$ and $\mathcal{L}=\{1,2,...,L\}$, respectively.
We assume each RIS can only be associated with one OP  at any given service duration \cite{TVT Cai}. In the $s$-th OP, there are $J_{s}$ BSs with $N$ antennas working in different sub-bands and providing service for $K_{s}$ single-antenna users. The set of all BSs and users in the $s$-th OP can be expressed by $\mathcal{J}_s =\{1,2,...,J_{s}\}$ and $\mathcal{K}_s = \{1,2,...,K_s\}$, respectively. Each user enjoys the benefits provided by all RISs associated with its belonging OP, and receives service from only one BS based on the OP's user association decision. This paper focuses on the joint optimization of RIS resources, user association, and beamforming design across multiple OPs. 

\subsection{Association Model}
In non-cooperative communication scenarios involving multiple OPs, RISs are provided and utilized as public resources that can be dynamically optimized to accommodate varying demands. Based on the available environmental information, the RP manages the association and reconfiguration of all $L$ RISs, i.e., determining which RISs serve specific OPs.
Specifically, different OPs are associated with respective RISs according to the association matrix designed by the RP, which can be denoted as $\mathbf{B} \triangleq [b_1,b_2,...,b_L]^T \in \mathbb{C}^{L\times 1}$, $b_l$ indicates the allocation of the $l$-th RIS and satisfies $b_l \in \mathcal{S}$. For the $s$-th OP, the set of its associated RISs can be given as
\begin{equation}
\label{eq_L}
\mathcal{L}_s = \big\{l \in \mathcal{L} \mid b_l = s \big\}, \forall s \in \mathcal{S},
\end{equation}
where $\bigcup_{s=1}^{S}\mathcal{L}_{s}=\mathcal{L}$. We define an element in $\mathcal{L}_{s}$ as $l_{s}$ and the size of $\mathcal{L}_{s}$ as $L_s$, satisfying $\sum_{s=1}^{S} L_s=L$.

To fully leverage the performance gains of spatial diversity, each OP should assign the most suitable BS to its subscribers based on their channel conditions. For the $s$-th OP with $J_s$ deployed BSs, user association design ensures more efficient resource utilization and broader coverage. The BS-user association  is denoted as $\mathbf{A}_{s} \triangleq [a_{1_s},a_{2_s},..., a_{K_s}]^T \in \mathbb{C}^{K_s \times 1}$, $a_{k_s}$ specifies the BS from which the $k_s$-th user receives service and satisfies $a_{k_s} \in \mathcal{J}_s, \forall s \in \mathcal{S}$. Similarly, for the $j_s$-th BS, the set of its associated users is expressed by
\begin{equation}
 \mathcal{K}_{j_s} = \big\{k_s \in \mathcal{K}_s \mid a_{k_s}= j_s \big\}, \forall j_s \in \mathcal{J}_s, \forall s \in \mathcal{S},
\end{equation}
which should satisfy $\bigcup_{{j_s}=1}^{J_s}\mathcal{K}_{j_s}=\mathcal{K}_{s}, \forall s \in \mathcal{S}$. We define an element in $\mathcal{K}_{j_s}$ as $k_{j_s}$ and the cardinality of $\mathcal{K}_{j_s}$ as $K_{j_s}$, satisfying $\sum_{j_s=1}^{J_s} K_{j_s}=K_s$.

\vspace{-0.1cm}
\subsection{RIS Model}
According to the practical reflection model, the equivalent impedance of the RIS can be adjusted by tuning the variable capacitance of its parallel resonant circuit, enabling precise control of the phase-shift within a specific frequency band \cite{CL Cai}, \cite{TCOM Li}. In practical mobile networks, OPs operate within separate and non-overlapping frequency bands, implying that the tunable capacitance ranges of different OPs are likewise nearly non-overlapping. In other words, the phase-shift response of RISs to signals from non-associated OPs is extremely limited and can often be regarded as negligible. Therefore, we adopt the simplified reflection model to characterize the frequency-selective properties of RIS\footnote{Specifically, the authors in \cite{TCOM Cai} demonstrate that the simplified model can achieve effective resource optimization with significantly reduced computational complexity. Moreover, our proposed algorithm can be seamlessly implemented within the practical reflection model.} \cite{TVT Liu}, \cite{TCOM Cai},  and \cite{TVT Cai}.

As shown in Fig. \ref{fig_system}(b) and Fig. \ref{fig_system}(c), each OP controls its corresponding RIS and designs the phase-shift matrix based on instructions issued by the RP. According to the simplified reflection model, the reflective elements of a given RIS provide adjustable phase-shifts only for signals within the frequency bands of its associated OP. For signals within the frequency bands of non-associated OPs, the RIS applies a fixed phase-shift of $0$/$2\pi$. Therefore, for the $s$-th OP, the phase-shift matrix of the $l$-th RIS in the environment can be expressed as
\begin{equation}
\label{eq_Phi}
\mathbf{\Phi}_{l} = \begin{cases} \mathbf{\Phi}_{l_s}, &  \mathrm{if} \ \exists l_s \in \mathcal{L}_s \ \mathrm{such \ that} \  l_s = l , \\
\mathbf{I}_{M_l}, & \mathrm{if} \ l_s \neq l  \ \mathrm{for \ all} \ l_s  \in \mathcal{L}_s, \end{cases}
\end{equation}
where $\mathbf{\Phi}_{l_s}= \mathrm{diag}\{e^{j\phi_{1_{l(s)}}},e^{j\phi_{2_{l(s)}}},...,e^{j\phi_{M_{l(s)}}}\}\in \mathbb{C}^{M_l \times M_l}$ represents the phase-shift matrix of the \(l\)-th RIS when controlled by the \(s\)-th OP, and $\phi_{m_{l(s)}} \in [0,2\pi], \forall  l\in \mathcal{L}$ denotes the adjustable phase-shift of the $m_l$-th element.

\vspace{-0.1cm}

\begin{figure}[t]
\centering
\includegraphics[width = 3.4in]{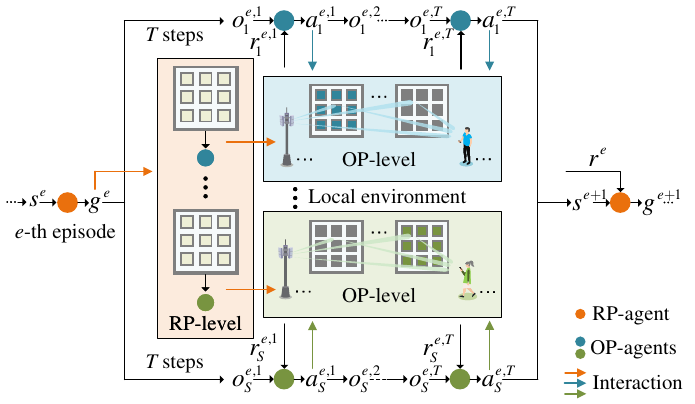}
\caption{HDRL framework for RP-OP coexisting system.}
\label{fig_HDRL}
%\vspace{-0.3 cm}
\end{figure}

\subsection{Channel Model}
For the $s$-th OP, we denote the channels from the $j_{s}$-th BS to the $k_{s}$-th user, from the $j_{s}$-th BS to the $l$-th RIS and from the $l$-th RIS to the $k_{s}$-th user as $\mathbf{h}^{\mathrm{d}}_{j_{s}, k_{s}} \in \mathbb{C}^{N\times 1}$, $\mathbf{G}_{j_{s}, l} \in \mathbb{C}^{M_l \times N}$ and $\mathbf{h}^{\mathrm{r}}_{l, k_{s}} \in \mathbb{C}^{M_l \times 1}$, respectively. All sub-channels follow Rician fading, which consists of deterministic line-of-sight (LoS) components and random non-line-of-sight (NLoS) components, expressed as
\begin{equation}
\begin{aligned}
\label{eq_subchannel}
\!&\mathbf{h}^{\mathrm{d}}_{j_{s}, k_{s}}=\sqrt{\eta(d^{\mathrm{BU}}_{j_{s}, k_{s}})}(\sqrt{\frac{\kappa}{\kappa+1}} \mathbf{h}^{\mathrm{d},\mathrm{LoS}}_{j_{s},k_{s}}
+\sqrt{\frac{1}{\kappa+1}} \mathbf{h}^{\mathrm{d},\mathrm{NLoS}}_{j_{s},k_{s}}), \\
\!& \mathbf{G}_{j_{s}, l}=\sqrt{\eta(d^{\mathrm{BR}}_{j_{s}, l})}(\sqrt{\frac{\kappa}{\kappa+1}} \mathbf{G}^{\mathrm{LoS}}_{j_{s}, l}+\sqrt{\frac{1}{\kappa+1}} \mathbf{G}^{\mathrm{NLoS}}_{j_{s}, l}), \\
\!&\mathbf{h}^{\mathrm{r}}_{l, k_{s}}=\sqrt{\eta(d^{\mathrm{RU}}_{l, k_{s}})}(\sqrt{\frac{\kappa}{\kappa+1}} \mathbf{h}^{\mathrm{r},\mathrm{LoS}}_{l, k_{s}}
+\sqrt{\frac{1}{\kappa+1}} \mathbf{h}^{\mathrm{r},\mathrm{NLoS}}_{l, k_{s}}), \\
\end{aligned}
\end{equation}
where $\kappa$ denotes the Rician factor, $d^{\mathrm{BU}}_{j_{s}, k_{s}}$, $d^{\mathrm{BR}}_{j_{s}, l}$ and $d^{\mathrm{RU}}_{l, k_{s}}$ denote the distances between the $j_s$-th BS and the $k_s$-th user, the $j_s$-th BS and the $l$-th RIS, and the $l$-th RIS and the $k_s$-th user, respectively. The distance-dependent channel path loss model $\eta(d)=C_0(\frac{d}{d_0})^{-\alpha}$, where $C_0$ represents the signal attenuation at the reference distance $d_0$ and $\alpha$ denotes the pass loss exponents. Therefore, the equivalent channel from the $j_{s}$-th BS to the $k_{s}$-th user can be written as
\begin{equation}
\label{eq_channel}
\mathbf{h}_{j_{s}, k_{s}} =  \sum_{\forall  l \in \mathcal{L}} \mathbf{G}_{j_{s}, l}^H \mathbf{\Phi}_{l} \mathbf{h}^{\mathrm{r}}_{l, k_{s}} + \mathbf{h}^{\mathrm{d}}_{j_{s}, k_{s}}, 
\end{equation}
where $\mathbf{\Phi}_{l}$ follows \eqref{eq_Phi} from the perspective of the $s$-th OP. For the $k_s$-th user, the matrix of all $J_s$ possible equivalent channel is denoted as $\mathbf{H}_{k_s}\triangleq [\mathbf{h}_{1_{s}, k_{s}},...,\mathbf{h}_{J_{s}, k_{s}} ] \in \mathbb{C}^{N \times J_s} $. The actual transmission channel for the $k_s$-th user is selected from $\mathbf{H}_{k_s}$ through user association matrix $\mathbf{A}_{s}$, expressed as
\begin{equation}
\mathbf{h}_{k_{s}} = \mathbf{h}_{j_{s}, k_{s}} \mid a_{k_s}= j_s, \forall j_{s} \in \mathcal{J}_s.
\end{equation}

For the $s$-th OP, the matrix of actual transmission channel for all $K_s$ users is denoted as $\mathbf{H}_{s}\triangleq [\mathbf{h}_{1_{s}},..., \mathbf{h}_{K_{s}}] \in \mathbb{C}^{N \times K_{s}}$.

\vspace{-0.2cm}

\subsection{Communication Model}
We consider the downlink communication of OP with all its subscribed users and assume that the frequency bands of all its BSs are non-overlapping. This effectively prevents the received signals from experiencing co-channel interference from other BSs. For the $s$-th OP, let $\mathbf{W}_{s}\triangleq [\mathbf{w}_{1_{s}},..., \mathbf{w}_{K_{s}}] \in \mathbb{C}^{N \times K_{s}}$ be the beamforming matrices for all its $J_s$ BSs. For the $j_s$-th BS,  $\mathbf{W}_{j_{s}}\in \mathbb{C}^{N \times K_{j_{s}}}$ is selected from $\mathbf{W}_{s}$ based on the corresponding BS-user association $\mathbf{A}_{s}$, by ensuring
\begin{equation}
\mathbf{W}_{j_{s}} = [\mathbf{w}_{k_s} \mid a_{k_s}= j_s, \mathbf{w}_{k_s} \in \mathbf{W}_{s}], \forall j_s \in \mathcal{J}_s, \forall s \in \mathcal{S},
\end{equation}
where a column in $\mathbf{W}_{j_{s}}$ is denoted as $\mathbf{w}_{k_{j_s}}, \forall k_{j_s} \in \mathcal{K}_{j_s}$. Therefore, the power constraint can be further expressed as
\begin{equation}
\sum_{\forall k_{j_s} \in \mathcal{K}_{j_s}} \| \mathbf{w}_{k_{j_s}} \|^{2} = P_{j_s}, \forall j_s, \forall s.
\end{equation}
where $P_{j_s}$ is the transmit power of the $j_s$-th BS and $P_s= \sum_{j_s=1}^{J_s}P_{j_s}$ represents the total power from all BSs corresponding to the $s$-th OP. With the beamforming matrix $\mathbf{W}_{j_{s}}$, all $J_s$ BSs in the $s$-th OP start the downlink communication with  $K_s$ users. The  received signal at the $k_{j_s}$-th user can be written as
\begin{equation}
\begin{aligned}
y_{k_{j_s}}=& \mathbf{h}_{j_{s}, k_{j_s}}^H \mathbf{w}_{k_{j_s}} x_{k_{j_s}} + \\
& \sum_{\bar{k}_{j_s} \neq k_{j_s},\forall \bar{k}_{j_s} \in \mathcal{K}_{j_s}} \mathbf{h}_{j_{s}, k_{j_s}}^H \mathbf{w}_{\bar{k}_{j_s}} x_{\bar{k}_{j_s}} +n_{j_{s}, k_{j_s}} ,
\end{aligned}
\end{equation}
where $x_{k_{j_s}}$ denotes the information symbol, $n_{j_{s}, k_{j_s}}\sim\mathcal{CN}(0, \sigma^{2})$ denotes the additive white Gaussian noise (AWGN). It is important to note that RIS association and phase-shifts directly influence the equivalent channel $\mathbf{h}_{j_{s}, k_{j_s}}$, while user association directly determines the set of served users $\mathcal{K}_{j_s}$.
Thus, the receive SINR at the $k_{j_s}$-th user can be calculated as
\begin{equation}
\gamma_{k_{j_s}}=\frac{\big|\mathbf{h}_{j_{s}, k_{j_s}}^H \mathbf{w}_{k_{j_s}}\big|^2}{\sum_{\bar{k}_{j_s} \neq k_{j_s},\forall \bar{k}_{j_s} \in \mathcal{K}_{j_s}} \big | \mathbf{h}_{j_{s}, k_{j_s}}^H \mathbf{w}_{\bar{k}_{j_s}} \big |^2+ \sigma^2}.
\end{equation}
Then, the sum-rate of all users within the $s$-th OP can be obtained as
\begin{equation}
R_{s} = \sum_{j_{s}=1}^{J_{s}} R_{j_s}=\sum_{j_{s}=1}^{J_{s}} \sum_{\forall k_{j_s} \in \mathcal{K}_{j_s}} \mathrm{log}_{2}(1+\gamma_{k_{j_s}}),
\end{equation}
where $R_{j_s}$ denotes the sum-rate of total $K_{j_s}$ users serving by the $j_s$-th BS.

\vspace{-0.1cm}

\subsection{Problem formulation}
From the RP's perspective, the optimization goal is to maximize the throughput of all $S$ OPs with the assistance of $L$ RISs. It can be observed that $\sum_{s=1}^{S} R_{s}$ is determined by the joint management of RIS allocation, BS-user association, BS transmit beamforming, and RIS passive beamforming.
Thus, this problem is formulated as
\begin{subequations}\label{eq:problem}
\begin{align}
\label{eq_problem}&\max _{\substack{ \mathbf{B}, \mathbf{A}_{s}, \mathbf{W}_{s}, \mathbf{\Phi}_{l}  \\  \forall l \in \mathcal{L}, \forall s \in \mathcal{S}}} \sum_{s=1}^{S} R_{s}\\ \label{eq_problem_b}
\text { s.t. } & 1 \leq b_l \leq S, \forall l,  \\ \label{eq_problem_L}
& \bigcup_{s=1}^{S}\mathcal{L}_{s}=\mathcal{L}, \\\ \label{eq_problem_a}
& 1 \leq a_{j_s} \leq J_s, \forall j_s, \forall s, \\ \label{eq_problem_K}
& \bigcup_{{j_s}=1}^{J_s}\mathcal{K}_{j_s}=\mathcal{K}_{s}, \forall s, \\ \label{eq_problem_Phi}
& \phi_{m_l} \in [0,2\pi], \forall m , \forall l , \\ \label{eq_problem_P}
& \sum_{\forall k_{j_s} \in \mathcal{K}_{j_s}} \| \mathbf{w}_{k_{j_s}} \|^{2} = P_{j_s}, \forall j_s, \forall s,
\end{align}
\end{subequations}
where \eqref{eq_problem_b}-\eqref{eq_problem_L} are constraints for RIS allocation matrix $\mathbf{B}$ ensuring each RIS is assigned to single OP, \eqref{eq_problem_a}-\eqref{eq_problem_K} are constraints for the user association matrix $\mathbf{A}_{s}, \forall s \in \mathcal{S}$ ensuring that each user within an OP is served by a corresponding BS, \eqref{eq_problem_Phi} is phase-shifts constraint for all RISs and \eqref{eq_problem_P} is the power constraint for each BS.

Unfortunately, directly deploying centralized joint optimization algorithms is impractical for \eqref{eq_problem}, primarily due to the inherent conflict between highly coupled optimization variables and entirely heterogeneous entities. For instance, the network topology is jointly determined by $\mathbf{B}$ and $\mathbf{A}_{s}$, which are designed by the RP and OPs, respectively. However, sharing critical operational information such as CSI for joint optimization is infeasible. This limitation exacerbates the already challenging association problem by introducing significant information insufficiency. Additionally, incorporating $\mathbf{W}_{s}$ and $\mathbf{\Phi}_{l}$ further increases the complexity of the joint resource optimization. Moreover, centralized optimization demands high-frequency interactions between the RP and OPs, which significantly undermines real-time operational efficiency. Therefore, we innovatively decompose \eqref{eq_problem} into sub-tasks executed separately by the RP and OPs. This enables different entities to independently optimize their internal resources based on the 
available information. In the next section, we will provide a detailed explanation of the proposed HDRL framework.

\section{Proposed Hierarchical DRL Approach}\label{section:3}

In this section, we utilize the SMDP theory to decompose \eqref{eq_problem} into forms that can be handled by RP and OPs, providing detailed definitions for each SMDP element to facilitate the interaction between RP and OPs. Then, based on the SMDP structure, we introduce a comprehensive HDRL framework to conduct resoure optimization.

\vspace{-0.3cm}
\subsection{SMDP Structure for RP-OP Coexisting Systems}
Converting \eqref{eq_problem} into the traditional MDP framework is not well-suited for RP-OP coexisting systems. On one hand, OPs can only observe environmental information related to their own service users and share only non-sensitive data with RP to facilitate communication. This makes it infeasible to centrally define states, actions, and rewards for both OPs and RP. On the other hand, RP should play a central role in multi-RIS-assisted systems, where it not only guides OPs in resource allocation through RIS scheduling, but also minimizes the overhead caused by frequent RIS switching between OPs. Therefore, we further refine the original MDP framework\footnote{Most policy optimization problems can be modeled as sequential decision models with Markov properties, represented by the MDP tuple $\langle\mathcal{S}, \mathcal{A},\mathcal{\pi}, \mathcal{R}, \mathcal{P}\rangle$. Specifically, the agent observes the environment's state $\mathcal{S}$, analyzes the current environmental information and combines it with policy $\mathcal{\pi}$ to take action $\mathcal{A}$. The agent then receives the reward $\mathcal{R}$ to evaluate the action and transitions to the next state based on the transition probabilities $\mathcal{P}$.}  by incorporating SMDP structure \cite{TCCN Zhou} and \cite{TCOM Geng}.

Firstly, to effectively deploy the SMDP structure into the multi-RIS multi-OP system, we introduce a two-time-scale procedure. Specifically, the real-time temporal sequence is abstracted into several episodes, each consisting of $T$ time steps, with the CSI remaining constant within each time step. Buliding on this, the RP and OPs can perform resource optimization at the episode and time-step scales, respectively. At each episode, RP designs the RIS allocation  to effectively schedule $L$ RISs to assist the communication services of $S$ OPs. The objective \eqref{eq_problem} for the RP can be decomposed into
\begin{equation}
\label{eq_problem_RP}
\max _{\substack{   \mathbf{B}, \mathbf{A}_{s}, \mathbf{W}_{s}, \mathbf{\Phi}_{l}   \\  \forall l \in \mathcal{L}, \forall s \in \mathcal{S}}} \sum_{s=1}^{S} R_{s}  \xrightarrow[\text{temporal abstraction}]{\text{RP decomposition}}
\max _{\substack{ \mathbf{B}^e}} \sum_{s=1}^{S} R_{s}^{e,t},
\end{equation}
where $\mathbf{B}^e$ denotes the RIS-OP association matrix at the $e$-th episode, $R_{s}^{e,t}$ represents the sum-rate of $K_s$ users for the $s$-th OP at the $t$-th time step of the $e$-th episode. Then, based on the RIS allocation in the current episode, each OP performs its respective resource allocation at each time step. Meanwhile, OP cannot share its instantaneous resource allocation design and QoS status with other OPs and RP. Therefore, for the $s$-th OP, the objective \eqref{eq_problem} is decomposed as
\begin{equation}
\label{eq_problem_OP}
\max _{\substack{ \mathbf{B}, \mathbf{A}_{s}, \mathbf{W}_{s}, \mathbf{\Phi}_{l}  \\  \forall l \in \mathcal{L}, \forall s \in \mathcal{S}}} \sum_{s=1}^{S} R_{s}  \xrightarrow[\text{temporal abstraction}]{\text{OP decomposition}}
\max _{\substack{ \mathbf{\Phi}_{l_{s}^e}^{e,t}, \mathbf{W}_{s}^{e,t}, \mathbf{A}_{s}^{e,t} \\  \forall l_s^e \in \mathcal{L}_s^e}}  R_{s}^{e,t},
\end{equation}
where $\mathcal{L}_s^e$ denotes the set of RISs associated with the $s$-th OP during the $e$-th episode, satisfying $\bigcup_{s=1}^{S}\mathcal{L}_{s}^{e}=\mathcal{L}$ as defined in \eqref{eq_L}. $\mathbf{W}_{s}^{e,t}$, $ \mathbf{A}_{s}^{e,t}$ denotes the beamforming matrix and user association matrix designed by the $s$-th OP at the $t$-th time step of the $e$-th episode, respectively.

Subsequently, we design the SMDP structure based on the two-time-scale procedure. As shown in Fig. \ref{fig_HDRL}, the proposed SMDP structure includes high-level transitions at the RP-level and sub-transitions at the OP-level, defined as follows
\begin{itemize}
\item RP-level: We regard RP as a single top-level agent responsible for managing all OPs, with its transition defined as $(s^e, g^e, r^e, s^{e+1})$. Here, $s^e$ represents the state of RP at the $e$-th episode, encompassing the shared information uploaded by all OPs. $g^e$ represents the goal output at the $e$-th episode, which can be viewed as a high-level action guiding the underlying agents in executing sub-policy planning. $r^e$ denotes the reward received by RP-agent at the end of the $e$-th episode. Unlike traditional MDPs, the transition from $s^e$ to $s^{e+1}$ unfolds over $T$ sub-time steps rather than occurring instantaneously.

\item OP-level: All $S$ OP-agents operate on an equal basis, performing their respective resource allocations based on the goal $g^e$ issued by RP-agent. At the $t$-th time step of the $e$-th episode, the transition of the $s$-th OP-agent is defined as $(o^{e, t}_{s}, a^{e, t}_{s}, r^{e, t}_{s}, o^{e, t+1}_{s})$. $o^{e, t}_{s}$ denotes the observation made by the $s$-th OP-agent at the $t$-th time step of the $e$-th episode, which includes its observable state information and $g^e$. $a^{e, t}_{s}$ and $r^{e, t}_{s}$ represent the action taken by the $s$-th OP-agent based on $o^{e, t}_{s}$ and the corresponding reward, respectively.
\end{itemize}

In summary, at the start of the $e$-th episode, the top-level RP-agent formulates $g^e$ based on $s^e$. Then, all $S$ OP agents at the low-level carry out respective sub-policy across $T$ time steps during the $e$-th episode, guided by the received $g^e$. For the $s$-th OP-agent at the $t$-th time step, it selects $a^{e, t}_{s}$ to interact with the environment based on $o^{e, t}_{s}$ and receives $r^{e, t}_{s}$. Upon completing interactions over $T$ time steps, the low-level agents upload shareable information to RP, enabling it to process and obtain the feedback $r^e$ for the $e$-th episode and determine $s^{e+1}$ for the next episode. 

Finally, we provide a detailed design of all the key elements in SMDP structure. It is crucial to prevent OPs from uploading sensitive information to RP while ensuring that all transitions remain feasible with manageable overhead. We start by defining RP-level elements $s^e$, $g^e$, and $r^e$ as follows.

\begin{itemize}
\item \textbf{RP-level state}: Designing RP's state presents significant challenges, as it must avoid directly collecting sensitive information from individual OPs, such as transmission data and CSI, while also preventing the upload of large data volumes, such as model information. To address this, we define an equivalent information matrix calculated by OP \cite{WCL Huang}. For the $s$-th OP, the information matrix at the $t$-th time step of the $e$-th episode is given as
\begin{equation}
\label{eq_information_matrix}
\hspace{-0.6 cm} \mathbf{U}^{e,t}_{s} = (\mathbf{H}^{e,t}_{s})^H \mathbf{W}^{e,t}_{s} \in \mathbb{C}^{K_{s} \times K_{s}}, \forall e, \forall t \in \{1, ..., T\},
\end{equation}
    where $\mathbf{H}^{e,t}_{s} \triangleq [\mathbf{h}^{e,t}_{1_{s}},..., \mathbf{h}^{e,t}_{K_{s}}]$ is equivalent transmission channel matrix of the $s$-th OP at the $t$-th time step of the $e$-th episode. $\mathbf{U}^{e,t}_{s}$ implicitly contains information about the OP's sensitive information, with its dimensions depending only on the number of service users. Accordingly, the RP's state can be defined as
\begin{equation}
\label{eq_RP_state}
\hspace{-0.0 cm} s^{e}  = \big\{ \mathbf{U}^{e-1}_{1},\mathbf{U}^{e-1}_{2},..., \mathbf{U}^{e-1}_{S} \big\},
\end{equation}
where
 \begin{equation}
\begin{aligned}\label{eq_OPs_upload}
\hspace{-0.0 cm} \mathbf{U}^{e}_{s} = \Big \{ \Re \big( \mathbf{U}^{e,1}_{s} \big),\Im \big( \mathbf{U}^{e,1}_{s} \big),..., \Re \big( \mathbf{U}^{e,T}_{s} \big),\Im \big( \mathbf{U}^{e,T}_{s} \big) \Big \}
\end{aligned}
\end{equation}
includes the environmental information for all $T$ time steps in the $e$-th episode for the $s$-th OP, $\Re(\cdot)$ and $\Im(\cdot)$ represent the real and imaginary parts, respectively.

\item \textbf{RP-level goal}: In the $e$-th episode, RP-agent outputs a scheduling scheme for $L$ RISs to guide $S$ OPs in resource allocation across the $T$ time steps of the episode. The RP-level goal is represented by
\begin{equation}
g^{e}  = \mathbf{B}^e \triangleq [b_1^e,b_2^e,...,b_L^e]^T,
\end{equation}
where $b_l^e \in \mathcal{S}$ represents the affiliation of the $l$-th RIS at the $e$-th episode.

\item \textbf{RP-level reward}: RP's objective is to maximize the sum-rate of all OPs in each episode by implementing an optimal RIS allocation scheme \eqref{eq_problem_RP}. Therefore, RP-level reward function is defined as
\begin{equation}
\label{eq_RP_reward}
r^{e}  = \frac{1}{T}\sum_{s=1}^{S} \sum_{t=1}^{T} R_{s}^{e,t}.
\end{equation}
\end{itemize}

Next, we define the key OP-level elements. For the $s$-th OP, $o^{e, t}_{s}$, $a^{e, t}_{s}$, and $r^{e, t}_{s}$ are designed as follows

\begin{itemize}

\item \textbf{OP-level observation}:
    OP-level observation includes RP-level goal for the given episode and the environmental state at the current time step. In this paper, we utilize the environmental information from the previous time step, thereby avoiding additional time cost of measuring the current environment. Additionally, given the high overhead and complexity required to precisely determine each sub-channel in the time-varying scenario, we assume that each OP can only accurately estimate the equivalent channel $\mathbf{h}_{j{s}, k_{s}}^{e,t}$ given in \eqref{eq_channel}. In summary, OP-level observation is defined as
\begin{equation}
\begin{aligned}\label{eq_OP_observation}
\hspace{-0.6 cm}o^{e,t}_{s}  = \Big \{ & \Re\big( \mathbf{H}_{1_s}^{e,t-1}\big), \Im\big( \mathbf{H}_{1_s}^{e,t-1}\big),...,  \\ & \Re\big( \mathbf{H}_{K_s}^{e,t-1}\big), \Im\big( \mathbf{H}_{K_s}^{e,t-1}\big), \mathbf{A}_{s}^{e,t-1}, g^e \Big \},
\end{aligned}
\end{equation}
where $\mathbf{H}_{k_s}^{e, t} \triangleq [\mathbf{h}_{1_{s}, k_{s}}^{e, t},...,\mathbf{h}_{J_{s}, k_{s}}^{e, t} ]$ is the matrix of equivalent channel from all $J_s$ BSs to the $k_s$-th user.

\item \textbf{OP-level action}:
Each OP further allocates resources based on RP-level goal for the current episode, including user association design, beamforming design, and RIS phase-shift design. The OP-level action is defined as
\begin{equation}
\begin{aligned}\label{eq_OP_action}
a^{e,t}_{s}  = \Big \{ &\phi_{1_{1(s)}}^{e,t},...,\phi_{M_{1(s)}}^{e,t},...,\phi_{1_{L(s)}}^{e,t},...,\phi_{M_{L(s)}}^{e,t}, \\
&\mathbf{A}_{s}^{e,t}, \Re\big(\mathbf{W}^{e,t}_{s} \big), \Im\big(\mathbf{W}^{e,t}_{s} \big) \Big \}.
\end{aligned}
\end{equation}

Although each OP designs the phase-shifts for all $L$ RISs, it only controls RISs specified by $g_e$. Therefore, from the perspective of the $s$-th OP, the phase-shift of the $l$-th RIS at the current time step is represented as
\begin{equation}\label{eq_OP_phase_shift}
\mathbf{\Phi}_{l}^{e,t} = \begin{cases} \mathrm{diag}\{e^{j\phi^{e,t} _{1_{l(s)}}},...,e^{j\phi^{e,t} _{M_{l(s)}}}\}, &  \mathrm{if} \ b_l^e= s, \\
\mathbf{I}_{M_l}, & \mathrm{else.} \  \end{cases}
\end{equation}

\item \textbf{OP-level reward}:
Each OP attempts to maximize the sum of the transmission rates for all its served $K_s$ users \eqref{eq_problem_OP}. Thus, the OP-level reward is expressed as
\begin{equation}\label{eq_OP_reward}
r^{e,t}_{s}  = R_{s}^{e,t}.
\end{equation}
\end{itemize}

\subsection{Proposed HDRL Framework based on SMDP}
Based on the above analysis, we propose an HDRL framework where each OP performs joint design of RIS phase-shifts, beamforming, and user association based on the RIS allocation scheme provided by the RP. Based on the HDRL framework, the complete resource optimization and interaction process between the RP and OPs, is summarized in Algorithm \ref{alg:1} and detailed as follows.

\begin{algorithm}[t]
\footnotesize
\caption{Proposed HDRL framework}
\label{alg:1}
\begin{algorithmic}[1]
\STATE{Initialize RP-agent and $S$ OP-agents.}\label{0-1}
\STATE{Initialize transitions buffers $B$, $B_1,...,B_S$.}\label{0-2}
    \FOR {each episode $e=0,1,...$}\label{0-3}
        \STATE {RP outputs $g^e$ through $s^e$.}\label{0-4}
        \FOR {each time step $t=1,...,T$}\label{0-5}
            \FOR {each OP $s=1,...,S$}\label{0-6}
            \STATE{Output $a_{s}^{e,t}$ \eqref{eq_OP_action} through $o_{s}^{e,t}$.}\label{0-7}
            \STATE{Control assigned RISs \eqref{eq_OP_phase_shift}, associate $K_s$ users through $\mathbf{A}_{s}^{e,t}$ and design beamforming $\mathbf{W}^{e,t}_{s}$.}\label{0-8}
            \STATE{Finish data transmission and receive $r_s^{e,t}$ \eqref{eq_OP_reward}.}\label{0-9}
            \STATE{Obtain $\mathbf{H}_{1_s}^{e,t},...,\mathbf{H}_{K_s}^{e,t}$ and form $o_{s}^{e,t+1}$ \eqref{eq_OP_observation}.}\label{0-10}
            \STATE{Calculate and store $\mathbf{U}^{e,t}_{s} = (\mathbf{H}^{e,t}_{s})^H \mathbf{W}^{e,t}_{s}$ \eqref{eq_information_matrix}.} \label{0-11}
            \STATE{Put transition $(o^{e, t}_{s}, a^{e, t}_{s}, r^{e, t}_{s}, o^{e, t+1}_{s})$ into $B_s$.}\label{0-12}
            \STATE{Optimize OP-level policy: Jump to Algorithm \ref{alg:3}.}\label{0-13}
            \IF{$t=T$}\label{0-15}
                \STATE{Upload $\mathbf{U}^{e}_{s}$ \eqref{eq_RP_state} and $\frac{1}{T} \sum_{t=1}^{T} R_{s}^{e,t}$ to RP.}\label{0-16}
            \ENDIF\label{0-17}
            \ENDFOR\label{0-14}
        \ENDFOR\label{0-18}
        \STATE {After receiving $\frac{1}{T} \sum_{t=1}^{T} R_{1}^{e,t}, ...,\frac{1}{T} \sum_{t=1}^{T} R_{S}^{e,t}$ and $\big\{ \mathbf{U}^{e}_{1},..., \mathbf{U}^{e}_{S} \big\}$
        from $S$ OPs, RP calculates $r^{e}$ \eqref{eq_RP_reward} and obtains $s^{e+1}$ \eqref{eq_RP_state}.}\label{0-19}
        \STATE{RP puts transition $(s^e, g^e, r^e, s^{e+1})$ into $B$.}\label{0-20}
        \STATE{Optimize RP-level policy: Jump to Algorithm \ref{alg:2}.}\label{0-21}
    \ENDFOR\label{0-22}
    \end{algorithmic}
\end{algorithm}

We abstract real-time temporal sequence into high-level episode and low-level time step, with each episode containing $T$ time steps. At the beginning of the $e$-th episode, RP-agent makes decisions based on $s^e$ and outputs $g^e$, determining RIS allocation for current episode in step \ref{0-4}.

For the $t$-th time step within the $e$-th episode, each OP independently allocates resources based on $g^e$, as outlined in steps \ref{0-7}-\ref{0-9}. In step \ref{0-7}, each OP-agent generates $a^{e,t}_{s}$ based on $o^{e,t}_{s}$, which includes beamforming for the $J_s$ BSs, user association for $K_s$ users, and phase-shift settings for $L$ RISs. In step \ref{0-8}, each OP controls the phase-shifts of its assigned RISs according to the RP-level goal $g^e$ and initiates the user data transmission process. Upon completion of the transmission, the OP-agent receives environmental feedback $r^{e,t}_{s}$ in step \ref{0-9}. After estimating the current equivalent channel, the OP-agent combines it with $g_e$ and $\mathbf{A}_{s}^{e,t}$ to form the environmental observation $o^{e,t+1}_{s}$ for the next time step in step \ref{0-10}. In step \ref{0-12}, the OP-agent stores $(o^{e, t}_{s}, a^{e, t}_{s}, r^{e, t}_{s}, o^{e, t+1}_{s})$ in its transition buffer $B_s$ for OP-level policy optimization, and then proceeds to the next time step.

After completing $T$ time steps, each OP-agent uploads the episode's average sum-rate $\frac{1}{T} \sum_{t=1}^{T} R_{s}^{e,t}$ and equivalent information matrix $\big\{ \mathbf{U}^{e}_{1},..., \mathbf{U}^{e}_{S} \big\}$ to RP in step \ref{0-16}. After receiving environmental information from all OPs, RP-agent processes the data in step \ref{0-19} to obtain $r^e$ and $s^{e+1}$, then stores $(s^e, g^e, r^e, s^{e+1})$ in its transition buffer $B$ in step \ref{0-20} for RP-level policy updates.

In the proposed hierarchical framework, OP-agents  extracts features from equivalent CSI and optimize respective resources. Meanwhile, RP-agent allocates RIS based on information matrix uploaded by all OPs. In the next section, we deploy HPPO algorithm to achieve policy optimization for all OPs. Notably, our proposed framework enables RP-agent to handle information from different OPs and deploy various algorithms for RIS allocation. A comprehensive comparison of these methods is provided in Sec. \ref{section:5}.

\section{Proposed HPPO Algorithm based on the hierarchical framework}\label{section:4}
In this section, based on HDRL framework, we deploy a HPPO algorithm to update network parameters for both RP and OPs. Our motivation for adopting PPO among various model-free DRL algorithms is well-founded. At the OP-level, PPO excels in real-time settings due to its on-policy nature, which enables rapid adaptation to highly stochastic and unpredictable environments, resulting in greater stability and reduced fluctuations. At the RP-level, value/policy-based algorithms struggle to balance exploration and exploitation, resulting in ineffective RIS allocation. Therefore, PPO algorithm, with its actor-critic framework that overcomes above shortcomings, also stands out as a strong contender. Additionally, to address the potential issue of output dimensionality explosion in large-scale RIS allocation, we propose an improved solution at the RP-level.

\subsection{HPPO Algorithm at the RP-Level}
RP-agent needs to process the OP information matrices from all time steps of previous episodes to allocate RISs. For effectively capturing both short-term and long-term temporal dependencies within each $\mathbf{U}^{e,t}_{s}$ \eqref{eq_OPs_upload}, long short-term memory (LSTM) networks are utilized to extract features from input state for RIS allocation.
Therefore, the actor and critic network parameters of RP-agent are denoted as $\theta=\{ \theta_{\mathrm{LSTM}}, \theta_{\mathrm{FC}} \}$ and $\psi=\{ \psi_{\mathrm{LSTM}}, \psi_{\mathrm{FC}} \}$, respectively.

The actor network processes $s^e$ sequentially through LSTM and fully-connected (FC) layers to derive real-time RIS allocation. Specifically, in the $e$-th episode, for the system with $S$ OPs assisted by $L$ RISs, the actor network outputs the probabilities for all $S^L$ possible allocation schemes and samples $g^e$ from goal policy $\pi_{\theta}(g \mid s^e)$. During interactions with the environment via the actor network, RP-agent stores transitions $(s^e, g^e, r^e, s^{e+1})$ in its buffer $B$ and updates old policy at intervals of every $E$ episodes.

\begin{figure*}[t]
\centering
\includegraphics[width = 7in]{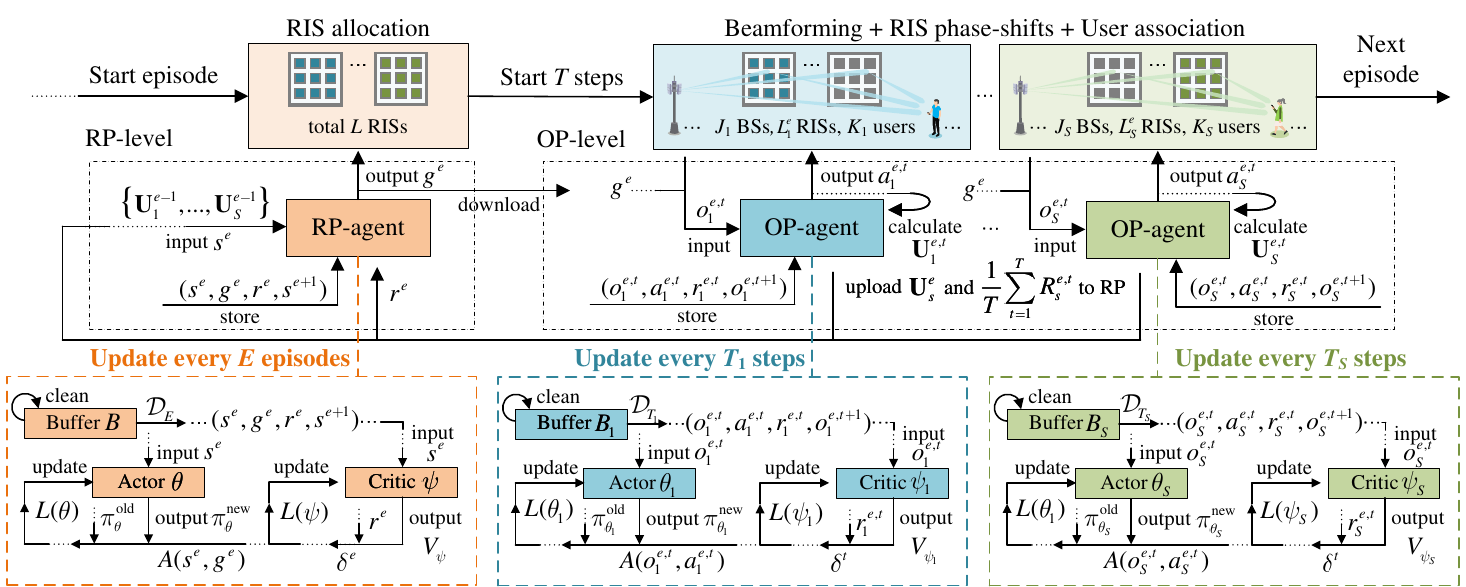}
\caption{Proposed HPPO algorithm in multi-RIS multi-OP networks.}
\label{fig_HPPO}
%\vspace{-0.3 cm}
\end{figure*}

RP-agent aims to obtain the optimal policy by updating the parameters of the actor network to maximize
\begin{equation}
J(\pi_{\theta})=\mathbb{E}_{\tau \sim \pi_{\theta}}\big[\sum_{e=0}^{\infty} (\gamma)^e r^e\big] \approx \mathbb{E}_{s^0 \sim \rho^0}\big[V_{\psi}(s^0)\big],
\end{equation}
where $\gamma$ is discount factor, $J(\pi_{\theta})$ denotes the expected return under policy $\pi_{\theta}$, which can be estimated by state-value $V_{\psi}(s^e)$ output by the critic network. According to \cite{NeurIPS Queeney}, in each update iteration using batch $\mathcal{D}_E$ of $E$ transitions, the new policy $\pi_{\theta}^{\text{new}}$ and the old policy $\pi_{\theta}^{\text{old}}$ satisfy
\begin{equation}
\begin{aligned}
\hspace{-0.2cm}& J(\pi_{\theta}^{\text {new }})-J(\pi_{\theta}^{\text {old }}) \geq \\
\hspace{-0.2cm}& \underbrace{\alpha{\mathbb{E}}_{\mathcal{D}_E}\big[\frac{\pi_{\theta}^{\text{new}}(g^e \mid s^e)}{\pi_{\theta}^{\text {old}}(g^e \mid s^e)} A(s^e, g^e)\big]}_{\text{surrogate objective (SO)}}- \underbrace{\beta{\mathbb{E}}_{\mathcal{D}_E}\Big[\big|\frac{\pi_{\theta}^{\text {new }}(g^e \mid s^e)}{\pi_{\theta}^{\text {old }}(g^e \mid s^e)}-1\big| \Big]}_{\text{penalty term (PT)}},
\end{aligned}
\end{equation}
for PPO algorithm, it is crucial to optimize network parameters to improve SO while constraining the difference between $\pi_{\theta}^{\text{new}}$ and $\pi_{\theta}^{\text{old}}$ within a certain range to prevent policy degradation caused by exponentially increasing PT values. Therefore, the optimization objective can be further expressed as
\begin{subequations}
\begin{align}
\label{p25a} & \mathop{\max}\limits_{\theta} ~ {\mathbb{E}}_{\mathcal{D}_E} \big[  \frac{\pi^{\text{new}}_{\theta}(g^e \mid s^e)}{\pi_{\theta}^{\text{old}}(g^{e} \mid s^{e})} A(s^e, g^e) \big], \\ \label{eq_HPPO_problem}
&~\text { s.t. } ~{\mathbb{E}}_{\mathcal{D}_E}\Big[\big|\frac{\pi_{\theta}^{\text {new }}(g^e \mid s^e)}{\pi_{\theta}^{\text {old }}(g^e \mid s^e)}-1\big| \Big] \leq \epsilon.
\end{align}
\end{subequations}
Therefore, RP-agent can constrain the update step size by clipping the policy ratio, ensuring that each RIS association strategy optimization proceeds positively in a stochastic environment. The loss function of actor network is defined as
\begin{equation}\label{eq_HPPO_RP_actor_loss}
\hspace{-0.3cm} L(\theta)=\mathbb{E}_{\mathcal{D}_E}\big[\min (\frac{\pi_{\theta}^{\text{new}}(g^e \mid s^e)}{\pi_{\theta}^{\text {old}}(g^e \mid s^e)}  A(s^e, g^e),  u (A(s^e, g^e)) )\big],
\end{equation}
where $u(\cdot)$ is used to satisfy the constraint \eqref{eq_HPPO_problem}, which is given by
\begin{equation}
u(A)= \begin{cases}(1+\epsilon) A , &  \mathrm{if} \ A \geq 0, \\ (1-\epsilon) A , & \mathrm{if} \ A<0.\end{cases}
\end{equation}

In PPO algorithm, advantage function $A(s^e, g^e)$ is used to evaluate whether the current choice of goal outperforms the average policy, thereby guiding the policy update. In this paper, we construct the advantage function using generalized advantage estimation (GAE), expressed as
\begin{equation}\label{eq_HPPO_RP_advantage}
A(s^e, g^e) = \delta^e+(\gamma \lambda) \delta^{e+1}+...+(\gamma \lambda)^{E-e+1} \delta^{E-1},
\end{equation}
where $\delta^{e} = r^{e} + \gamma V_{\psi}(s^{e+1})- V_{\psi}(s^{e})$ represents one-step temporal difference (TD). The critic network evaluates the state-action value $Q_{\psi}(s^e, g^e)$ by constructing the residual $r^{e} + \gamma V_{\psi}(s^{e+1})$, while subtracting the current state value $V_{\psi}(s^{e})$ to reduce estimation variance, thereby providing more accurate guidance to the actor. The loss function of critic network can be given as
\begin{equation}\label{eq_HPPO_RP_critic_loss}
L(\psi) = {\mathbb{E}}_{\mathcal{D}_E}\big[ r^{e} + \gamma  V_{\psi}(s^{e+1})- V_{\psi}(s^{e}) \big].
\end{equation}

\begin{algorithm}[t]
\footnotesize
\caption{{HPPO for updating the RP-agent}}
\label{alg:2}
\begin{algorithmic}[1]
\IF{collect $ E$ transitions}
    \STATE{Output $V_{\psi}(s^1),...,V_{\psi}(s^{E+1})$ through critic.}
    \STATE{Estimate $A(s^1,g^1),...,A(s^E,g^E)$ \eqref{eq_HPPO_RP_advantage}.}
    \STATE{Calculate actor's loss function $L(\theta)$ \eqref{eq_HPPO_RP_actor_loss}.}
    \STATE{Update actor parameters $\theta \longleftarrow \theta +\beta_{\theta} \cdot \nabla_{\theta} L(\theta)$.}
    \STATE{Calculate critic's loss function $L(\psi)$ \eqref{eq_HPPO_RP_critic_loss}.}
    \STATE{Update critic parameters $\psi \longleftarrow \psi +\beta_{\psi} \cdot \nabla_{\psi} L(\psi)$.}
    \STATE{Clear all transitions in $B$.}
\ENDIF
\end{algorithmic}
\end{algorithm}

Algorithm \ref{alg:2} presents the parameter update process using PPO. During each episode, RP-agent leverages the actor network to dynamically manage RIS allocation in real-time, and every $E$ episodes, it optimizes the allocation strategy based on stored transitions. In the next subsection, we will similarly apply PPO algorithm to update OP-agents.

\subsection{HPPO Algorithm at the OP-Level}

For the $s$-th OP-agent, we define the parameters of the actor and critic networks, composed of FC layers, as $\theta_c$ and $\psi_c$, respectively. Meanwhile, we use Beta distribution to fit policy $\pi_{\theta_c}$, avoiding potential performance degradation that can result from the additional clipping required for unbounded Gaussian-distributed outputs \cite{ICML Chou}. The estimated policy is written as
\begin{equation}
\pi_{\theta_c}(a \mid o_s^{e,t})=\frac{a^{\alpha(o_s^{e,t})-1}(1-a)^{\beta(o_s^{e,t})-1}}{B(\alpha(o_s^{e,t}), \beta(o_s^{e,t}))},
\end{equation}
where outputs $\alpha(o_s^{e,t})$ and $\beta(o_s^{e,t})$ define the shape of Beta distribution and $B(\alpha, \beta)=\int_0^1 t^{\alpha-1}(1-t)^{\beta-1} d t$. Therefore, the $s$-th OP-agent can derive the resource allocation scheme for the $t$-th time step in the $e$-th episode by reconstructing sampled action $a_s^{e,t}$.

At the OP-level, every $T_s$ time steps, the $s$-th OP agent updates its network parameters using $\mathcal{D}_{T_s}$ from transitions buffer $B_s$. Similar to the previous subsection, the advantage function for the $s$-th OP-agent is expressed as
\begin{equation}\label{eq_HPPO_OP_advantage}
A(o^{e,t}_{s}, a^{e,t}_{s}) = \delta^{t}+(\gamma \lambda) \delta^{t+1}+...+(\gamma \lambda)^{T_s-t+1} \delta^{T_s-1},
\end{equation}
where $\delta^{t} = r_{s}^{e, t} + \gamma V_{\psi_{s}}(o_{s}^{e, t+1})- V_{\psi_s}(o_{s}^{e, t})$. In the same way, the loss functions for $\theta_{s}$  and $\psi_{s}$ are respectively given as
\begin{subequations}
\begin{align}
\hspace{-0.3cm} &
\label{eq_HPPO_OP_actor_loss}L(\theta_s)= \\
&\mathbb{E}_{\mathcal{D}_{T_s}}\big[\min (\frac{\pi_{\theta_s}^{\text{new}}(a_{s}^{e,t} \mid o_{s}^{e,t})}{\pi_{\theta_s}^{\text {old}}(a_{s}^{e,t} \mid o_{s}^{e,t})}  A(a_{s}^{e,t}, o_{s}^{e,t}),  u (A(a_{s}^{e,t}, o_{s}^{e,t})) )\big],\nonumber \\
&\label{eq_HPPO_OP_critic_loss}L(\psi_s) = {\mathbb{E}}_{\mathcal{D}_{T_s}}\big[ r_{s}^{e, t} + \gamma  V_{\psi_s}(o_{s}^{e,t+1})- V_{\psi_s}(o_{s}^{e, t}) \big].
\end{align}
\end{subequations}

\begin{algorithm}[t]
\footnotesize
\caption{{HPPO for updating the $s$-th OP-agent}}
\label{alg:3}
\begin{algorithmic}[1]
\IF{collect $T_s$ transitions}
    \STATE{Output $V_{\psi_s}(o_{s}^{e,t}), \forall o_{s}^{e,t}\in \mathcal{D}_{T_s}$ through critic.}
    \STATE{Estimate $A(o^{e,t}_{s}, a^{e,t}_{s}), \forall (o^{e,t}_{s}, a^{e,t}_{s}) \in \mathcal{D}_{T_s}$ \eqref{eq_HPPO_OP_advantage}.}
    \STATE{Calculate actor's loss function $L(\theta_s)$ \eqref{eq_HPPO_OP_actor_loss}.}
    \STATE{Update actor parameters $\theta_s \longleftarrow \theta_s +\beta_{\theta_s} \cdot \nabla_{\theta_s} L(\theta_s)$.}
    \STATE{Calculate critic's loss function $L(\psi_s)$ \eqref{eq_HPPO_OP_critic_loss}.}
    \STATE{Update critic parameters $\psi_s \longleftarrow \psi_s +\beta_{\psi_s} \cdot \nabla_{\psi_s} L(\psi_s)$.}
    \STATE{Clear all transitions in $B_s$.}
\ENDIF
\end{algorithmic}
\end{algorithm}

In summary, our implemented HPPO algorithm facilitates asynchronous updates for RP and OPs within the hierarchical framework, ensuring real-time policy optimization. 

\begin{table*}[h]
    \centering
    \footnotesize
    \caption{Comparison between HPPO and S-HPPO at the RP-level.}
    \begin{tabular}{|c|c|c|}
    \hline
    Algorithm & Input/output width of actors & Computational complexity (inference)\\
    \hline
    HPPO & $H_{\mathrm{a}}^{0}=2\sum_{s=1}^{S}K_s^2, H_{\mathrm{a}}^{N_{\mathrm{a}}}=S^L$ & $\mathcal{O}\Big(T\sum_{n=0}^{N^{\mathrm{L}}_{\mathrm{a}}-1}H_{\mathrm{a}}^{n}H_{\mathrm{a}}^{n+1}+  \sum_{n=N^{\mathrm{L}}_{\mathrm{a}}}^{N_{\mathrm{a}}-1}H_{\mathrm{a}}^{n}H_{\mathrm{a}}^{n+1}\Big)$ \\
    % & Update & $\mathcal{O}\Big(T\sum_{n=0}^{N^{\mathrm{L}}_{\mathrm{a}}-1}H_{\mathrm{a}}^{n}H_{\mathrm{a}}^{n+1}+  \sum_{n=N^{\mathrm{L}}_{\mathrm{a}}}^{N_{\mathrm{a}}-1}H_{\mathrm{a}}^{n}H_{\mathrm{a}}^{n+1}\Big)$ \\
    \hline
    S-HPPO & $H_{\mathrm{a}}^{0_l}=l+2\sum_{s=1}^{S}K_s^2, H_{\mathrm{a}}^{{N_{\mathrm{a}}}_l}=S$ & $\mathcal{O}\Big( \sum_{l=0}^{L-1}\big(T\sum_{n=0}^{N^{\mathrm{L}}_{\mathrm{a}}-1}H_{\mathrm{a}}^{n_l}H_{\mathrm{a}}^{n+1_l} +  
\sum_{n=N^{\mathrm{L}}_{\mathrm{a}}}^{N_{\mathrm{a}}-1}H_{\mathrm{a}}^{n_l}H_{\mathrm{a}}^{n+1_l}\big) \Big)$ \\
    \hline
    \end{tabular}
    \label{table_complexity}
\end{table*}

\subsection{Improved S-HPPO Algorithm at the RP-Level for Large-Scale RIS Allocation}
In this subsection, we propose an effective solution to address the major challenge of deploying HPPO algorithm in multi-RIS multi-OP networks. In proposed HPPO algorithm, RP-agent outputs all possible discrete probability value to construct the real-time policy $\pi_\theta$. However, it is important to note that the dimensionality of the goal space is $S^L$, which grows exponentially with the number of OPs and RISs. In large-scale RIS allocation environments, the ultra-high-dimensional output space at the RP-level not only increases the risk of insufficient learning leading to local optima, but also raises hardware costs to maintain real-time updates and inference, further complicating the practical deployment of the HPPO algorithm. Therefore, we extend the approach proposed in \cite{Arxiv Metz} to further decompose the RP-level MDP, thereby reducing the dimensionality of the output space. The improvement scheme based on HPPO is named as sequential-HPPO (S-HPPO).

Considering that $\mathbf{B}^e$ can be equivalently viewed as a combination of $L$ sub-optimization variables $b_1^e, b_2^e, ..., b_L^e$, the intuition behind reducing the output space is to decompose it into $L$ $S$-dimensional sub-goal spaces. Simultaneously, this decomposition must be performed without altering the original MDP structure, meaning that the elements of all sub-MDPs should not depend on additional observations from the environment, but rather on the top-level MDP. Therefore, for the unchanged top-level transitions $(s^e, g^e, r^e, s^{e+1})$, the decomposed lower-level transitions can be defined as $(o^{e,l}, g^{e,l}, r^{e,l}), l=1,2,...,L$. For the $l$-th decomposed transition, $o^{e,l} = \{s^e, g^{e, 1},g^{e, 2},...,g^{e, l-1} \}$  include the original state and all past $l-1$ steps' output sub-goals and satisfy $o^{e,1} = s^e$, $g^{e,l} = b_l^e$ represents the allocation of the $l$-th RIS in the $l$-th step of the $e$-th episode. According to \cite{Arxiv Metz}, we provide the low-level external reward $r^{e,L}=r^e$, while the rewards for all other steps $l=1,2,...,L-1$ are set to 0.

\begin{figure}[t]
\centering
\includegraphics[width = 3.4in]{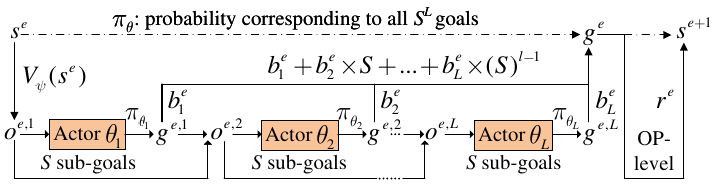}
\caption{Improved S-HPPO structure for large-scale goal space.}
\label{fig_S-HPPO}
%\vspace{-0.3 cm}
\end{figure}

Next, we introduce the S-HPPO algorithm for large-scale RIS allocation by integrating the hierarchical MDPs at the RP-level. As shown in Fig. \ref{fig_S-HPPO}, to model the lower-level sequential policy, we define $\mathcal{\pi} = [ \pi_{\theta_1},...,\pi_{\theta_L}]$ where each sub-policy $\pi_{\theta_l}$ is constructed by an actor network with $\theta_l$. In the $e$-th episode, the $l$-th actor receives observational information from past $l-1$ actors and samples its policy $\pi_{\theta_l}(g \mid s^{e,l})$ to determine the allocation of the $l$-th RIS. After all $L$ RISs have been sequentially decided, the RP-agent derives $g^{e}$ from $[g^{e, 1},...,g^{e, l}]$. By increasing the number of actors (MDP steps) from 1 to $L$, we reduce the dimensionality of the goal output space from $S^L$ to $S$, making it feasible to update model parameters in large-scale RIS environment.

In terms of policy updates, since the lower-level transitions do not incorporate additional environmental information, we can continue to use $\mathcal{D}_{E}$ to update actors. Additionally, to ensure consistency in the hierarchical MDP, we employ a critic network to fit the top-level MDP and simultaneously guide the lower-level actors. The critic network $\psi$ retains a high degree of consistency with the previous HPPO algorithm and continues to use the loss function \eqref{eq_HPPO_RP_critic_loss}. The loss function of the $l$-th actor network can be denoted as
\begin{equation}
 L(\theta_l)=\mathbb{E}_{\mathcal{D}_E}\big[\min (\frac{\pi_{\theta_l}^{\text{new}}(g^{e, l} \mid s^{e, l})}{\pi_{\theta_l}^{\text {old}}(g^{e, l} \mid s^{e, l})}  A(s^e, g^e),  u (A(s^e, g^e)) )\big],
\end{equation}
where $g^{e, l}$ and $s^{e, l}$ can be obtained through top-level transitions $(s^e, g^e, r^e, s^{e+1})$. The policy update process in S-HPPO algorithm is similar to Algorithm \ref{alg:2}, with the difference that the $l$-th actor updates its parameters using gradient ascent $\theta_l \longleftarrow \theta_l +\beta_{\theta_l} \cdot \nabla_{\theta_l} \sum_{l=1}^{L}L(\theta_l)$.

\subsection{Complexity Analysis}

In this subsection, we provide the computational complexity analysis for S-HPPO and HPPO algorithms.

Firstly, at the RP-level, we configure the HPPO with an actor-critic structure, where the actor consists of $N_{\mathrm{a}}$ layers and the critic consists of $N_{\mathrm{c}}$ layers. Specifically, the actor (critic) includes $N^{\mathrm{L}}_{\mathrm{a}}$ ($N^{\mathrm{L}}_{\mathrm{c}}$) layers of LSTM and $N^{\mathrm{F}}_{\mathrm{a}}$ ($N^{\mathrm{F}}_{\mathrm{c}}$) layers of FC , respectively. The input and output widths of the $n$-th layer of the actor(critic) are denoted as $H_{\mathrm{a}}^{n-1}(H_{\mathrm{c}}^{n-1})$ and $H_{\mathrm{a}}^{n}(H_{\mathrm{c}}^{n})$, respectively. Similarly, the $l$-th actor of the S-HPPO consists of $N_{\mathrm{a}}=N^{\mathrm{L}}_{\mathrm{a}}+N^{\mathrm{F}}_{\mathrm{a}}$ layers, with the input and output widths of the $n$-th layer denoted as $H_{\mathrm{a}}^{{n-1}_l}$ and $H_{\mathrm{a}}^{{n}_l}$, respectively.

Table \ref{table_complexity} compares the computational complexity of the inference phase for S-HPPO and HPPO. The analysis shows that, by further decomposing the top-level MDP, the improved S-HPPO algorithm provides an effective solution  in large-scale RIS allocation scenarios. For the update phase, the computational complexity at the RP-level for S-HPPO and HPPO is given as $\mathcal{O}\big(E(T\sum_{n=0}^{N^{\mathrm{L}}_{\mathrm{a}}-1}H_{\mathrm{a}}^{n}H_{\mathrm{a}}^{n+1}+\sum_{n=N^{\mathrm{L}}_{\mathrm{a}}}^{N_{\mathrm{a}}-1}H_{\mathrm{a}}^{n}H_{\mathrm{a}}^{n+1}+T$\\$\sum_{n=0}^{N^{\mathrm{L}}_{\mathrm{c}}-1}H_{\mathrm{c}}^{n}H_{\mathrm{c}}^{n+1}+\sum_{n=N^{\mathrm{L}}_{\mathrm{c}}}^{N_{\mathrm{c}}-1}H_{\mathrm{c}}^{n}H_{\mathrm{c}}^{n+1} )\big)$ and $\mathcal{O}\big(E(\sum_{l=0}^{L-1}(T$\\
$\sum_{n=0}^{N^{\mathrm{L}}_{\mathrm{a}}-1}H_{\mathrm{a}}^{n_l}H_{\mathrm{a}}^{n+1_l}+\sum_{n=N^{\mathrm{L}}_{\mathrm{a}}}^{N_{\mathrm{a}}-1}H_{\mathrm{a}}^{n_l}H_{\mathrm{a}}^{n+1_l})+T\sum_{n=0}^{N^{\mathrm{L}}_{\mathrm{c}}-1}H_{\mathrm{c}}^{n}$\\$H_{\mathrm{c}}^{n+1}+\sum_{n=N^{\mathrm{L}}_{\mathrm{c}}}^{N_{\mathrm{c}}-1}H_{\mathrm{c}}^{n}H_{\mathrm{c}}^{n+1} ) \big)$, respectively.

Furthermore, we analyze the OP-level's computational complexity, which is the same for both S-HPPO and HPPO. We assume that the actor (critic) networks of each OP consist of $N_{\mathrm{a}}$ ($N_{\mathrm{c}}$) layers of FC. For the $s$-th OP-agent, the input and output widths of the $n$-th layer of the actor (critic) are denoted as $H_a^{n-1_s}$ ($H_{\mathrm{c}}^{n-1_s}$) and $H_{\mathrm{a}}^{n_s}$ ($H_{\mathrm{c}}^{n_s}$), respectively. Therefore, for the $s$-th OP-agent, the computational complexities of inference and update are expressed as $\mathcal{O}\big(\sum_{n=0}^{N_{\mathrm{a}}-1}H_{\mathrm{a}}^{n_s}H_{\mathrm{a}}^{{n+1}_s}\big)$ and $\mathcal{O}\big(T_s(\sum_{n=0}^{N_{\mathrm{a}}-1}H_{\mathrm{a}}^{n_s}H_{\mathrm{a}}^{{n+1}_s} + \sum_{n=0}^{N_{\mathrm{c}}-1}H_{\mathrm{c}}^{n_s}H_{\mathrm{c}}^{{n+1}_s} )\big)$, respectively. In the next section, we will present a detailed comprehensive simulation.

\section{Simulation Results}\label{section:5}

\begin{table}[t]
    \centering
    \footnotesize
    \caption{Environment parameters and HPPO parameters}
    \label{table_parameters}
    \begin{tabular}{|c|c|c|c|}
    \hline
    Environment  & \multirow{2}{*}{Values} & HPPO  & \multirow{2}{*}{Values} \\
    Parameters & & Parameters & \\
    \hline
    $S$ & 2 & $\gamma$ & 0.99 \\
    \hline
    $J_s$ & 2 & $\lambda$ & 0.95 \\
    \hline
    $K_s$ & 5  & $T_s$ & 30 \\
    \hline
    $N$ & 10 & $E$ & 30 \\
    \hline
    $L$ & 4 & $\epsilon$ & 0.2 \\
    \hline
    $M_l$ & 20 & $\beta_{\psi}$ & $1 \times 10^{-4}$ \\
    \hline
    $\kappa$ & 10 & $\beta_{\phi}$ & $1 \times 10^{-4}$ \\
    \hline
    $P_{j_s}$ & 0.5W & $\beta_{\psi_s}$ & $2 \times 10^{-4}$ \\
    \hline
    $T$ & 10 & $\beta_{\phi_s}$ & $2 \times 10^{-4}$ \\
    \hline
    \end{tabular}
\end{table}

\begin{figure*}[t]
    \centering

    % First row of images
    \begin{minipage}{0.32\textwidth}
        \centering
        \includegraphics[width=\linewidth]{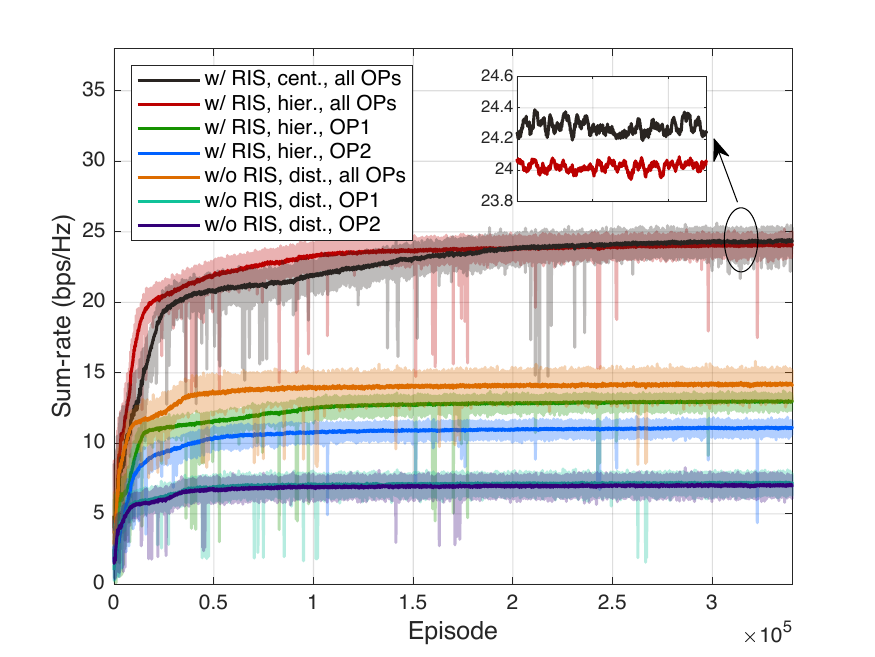}
        \caption{Sum-rate convergence for each OP.}
        \label{fig_curves_2OPs}
    \end{minipage}\hfill
    \begin{minipage}{0.32\textwidth}
        \centering
        \includegraphics[width=\linewidth]{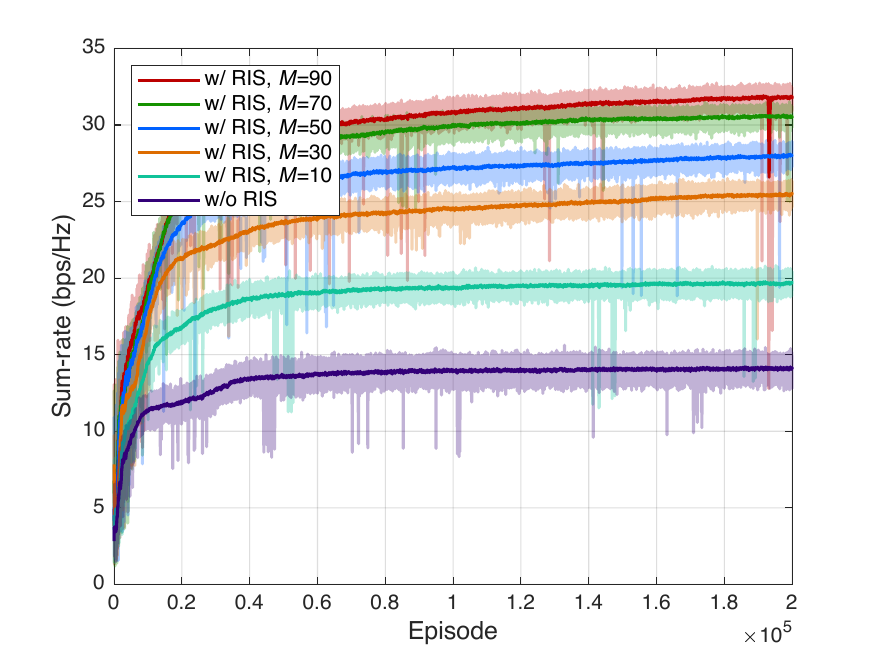}
        \caption{Sum-rate convergence for different RIS elements.}
        \label{fig_curves_M}
    \end{minipage}\hfill
    \begin{minipage}{0.32\textwidth}
        \centering
        \includegraphics[width=\linewidth]{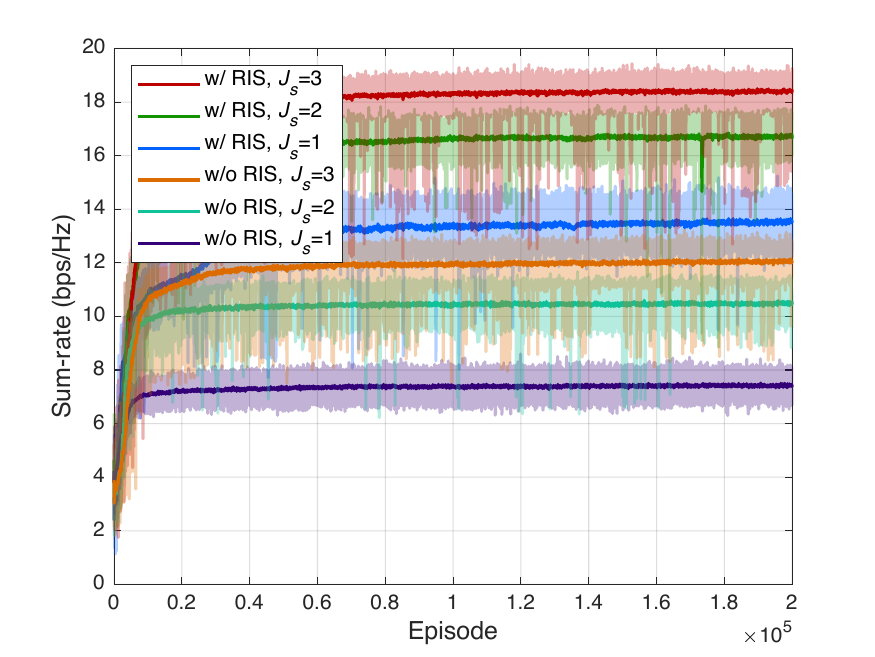}
        \caption{Sum-rate convergence for different number of BSs.}
        \label{fig_curves_J}
    \end{minipage}

    \vspace{1em}

    % Second row of images
    \begin{minipage}{0.32\textwidth}
        \centering
        \includegraphics[width=\linewidth]{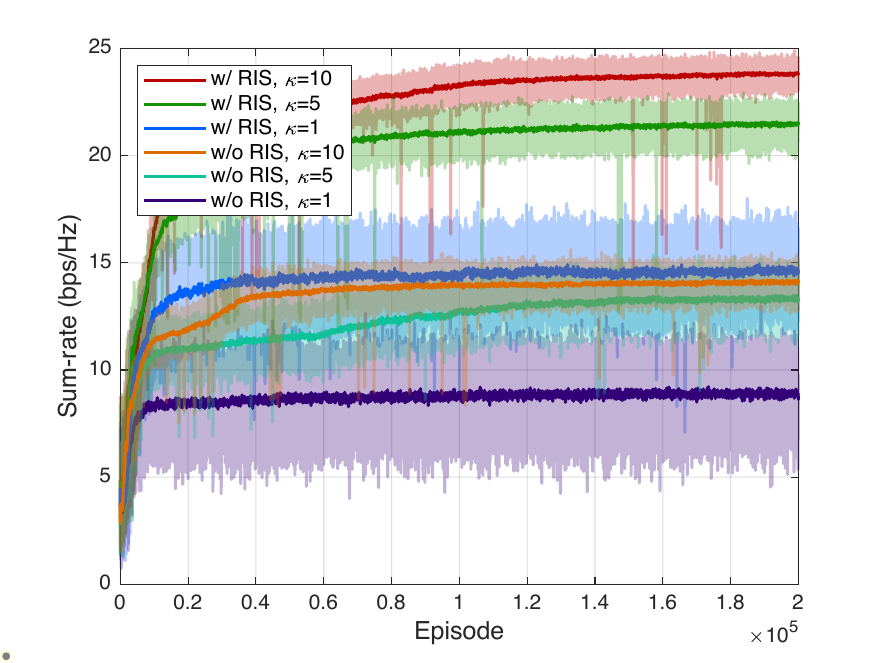}
        \caption{Sum-rate convergence for different Rician factor.}
        \label{fig_curves_rician}
    \end{minipage}\hfill
    \begin{minipage}{0.32\textwidth}
        \centering
        \includegraphics[width=\linewidth]{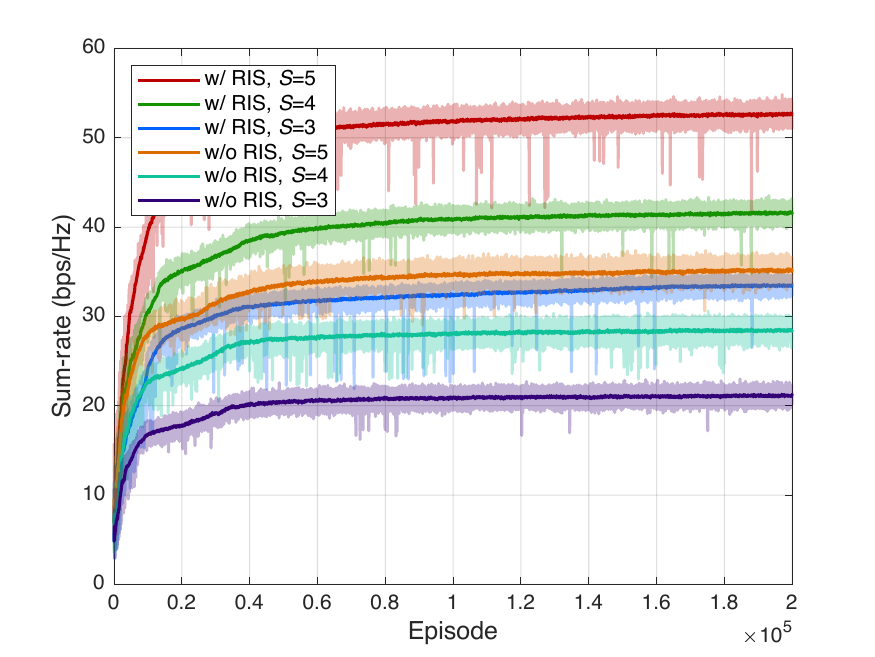}
        \caption{Sum-rate convergence for different number of OPs.}
        \label{fig_curves_S}
    \end{minipage}\hfill
    \begin{minipage}{0.32\textwidth}
        \centering
        \includegraphics[width=\linewidth]{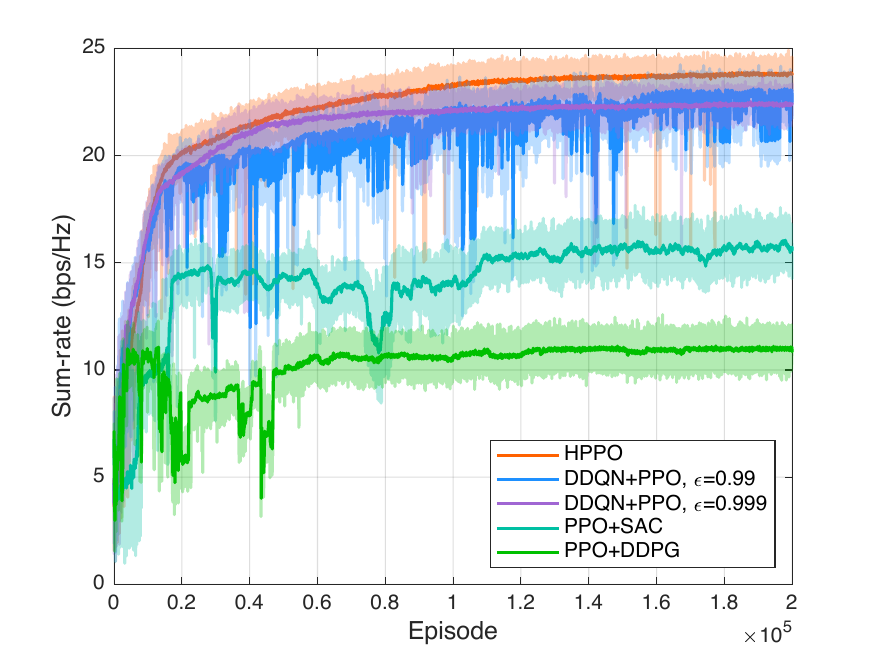}
        \caption{Sum-rate convergence for different DRL algorithms.}
        \label{fig_curves_DRL}
    \end{minipage}
\end{figure*}

In this section, we conduct extensive experiments for performance evaluation. Firstly, we verify the stability of the HPPO algorithm under various environmental settings. Secondly, we compare the performance of HPPO against various RIS allocation benchmarks based on the HDRL framework. Finally, we perform a detailed comparison between S-HPPO and HPPO in large-scale RIS allocation scenarios.

\subsection{Simulation Settings}
We consider a multi-RIS multi-OP network, where BSs are randomly distributed along a semicircular region with a radius of $150\text{m}$. The RISs are positioned along the x-axis between coordinates $(-5\text{m}, 0)$ and $(5\text{m}, 0)$, while users are located within a semi-circular annular region with inner and outer radii of $5\text{m}$ and $7\text{m}$, respectively.
The pass-loss exponents for the BS-RIS, RIS-user, and BS-user channels are 2.5, 2.8, and 3.5, respectively. In the proposed HPPO algorithm, each OP-agent is structured with an actor-critic network, both consisting of 4 FC layers. The RP-agent employs an actor-critic framework, incorporating 1 LSTM layer followed by 3 FC layers. All hidden layers across the networks have a width of 500.
%It is important to highlight that the DRL convergence plots are crucial in this scenario, as they capture the interactions and policy updates of each agent within the stochastic environment.
In this paper, instantaneous sum-rates   for each episode are depicted with transparent lines, while the smoothed average rate over a 100-episode window is shown with solid lines. All simulations are conducted using an NVIDIA GeForce GTX 1660 SUPER. Other default environment parameters are listed in Table \ref{table_parameters}.

\subsection{Stability of HPPO in Multi-RIS Multi-OP Networks}

Fig. \ref{fig_curves_2OPs} illustrates the sum-rate variation of each OP. On the one hand, we simulate the scenario without RIS (w/o RIS), where each OP uses the PPO algorithm in a \textbf{distributed} manner for beamforming and user association design. The results show that the RP’s allocation of RIS resources leads to a 85\% and 57\% sum-rate improvement for two OPs, respectively. This demonstrates that the proposed HPPO algorithm can achieve collaborative optimization of coupled resources between the RP and OPs. On the other hand, we compare an idealized \textbf{centralized} strategy, where a PPO-agent performs joint resource optimization based on the operational information of $S$ OPs. Experimental results demonstrate that both the  HPPO algorithm and the centralized optimization successfully learned the optimal RIS allocation for the RP. Although there is some performance loss at the OP-level, the results still validate the effectiveness of our HDRL framework. Specifically, OPs can extract key features from real-time environmental information, while the RP efficiently allocates RIS resources based on the implicit information provided by OPs. Moreover, HPPO exhibits faster convergence, proving that \textbf{hierarchical} task decomposition accelerates learning efficiency.

To further  assess  the stability of HPPO algorithm, Figs. \ref{fig_curves_M}-\ref{fig_curves_S} illustrate the sum-rate variation of all OPs ($\sum_{s=1}^{S}\sum_{t=1}^{T} R_{s}^{e,t}$) under different environmental parameters. Specifically, Fig. \ref{fig_curves_M} shows the effect of varying RIS elements $M_l$, demonstrating that OPs can effectively design phase-shifts based on the RISs allocated by the RP. 
In Fig. \ref{fig_curves_J}, we vary the number of BS $J_s$. The experimental results demonstrate that when $J_s = 2$, performance gains of 41\% and 22\% are achieved under w/ RIS and w/o RIS scenarios, respectively. This indicates that incorporating BS-user association can further leverage spatial diversity gains. Notably, when $J_s = 3$, these gains decrease to 14\% and 10\%, respectively, due to the proportional reduction in available bandwidth per BS.  Fig. \ref{fig_curves_rician} shows the effect of different \(\kappa\) on the HPPO algorithm. Although increased NLoS random factors in the environment make feature extraction more challenging for OPs, resulting in performance degradation and greater fluctuations, HPPO still provides stable sum-rate gains. Finally, Fig. \ref{fig_curves_S} shows the impact of changing the number of OPs $S$, proving that HPPO's distributed optimization of coupled resources is not limited by the number of agents. In summary, the proposed HPPO algorithm demonstrates excellent stability across various environmental parameters in multi-RIS multi-OP networks.

Additionally, in Fig. \ref{fig_curves_DRL}, we employ various model-free DRL algorithms based on the proposed HDRL framework. At the OP-level, we implement soft actor-critic (SAC) and deep deterministic policy gradient (DDPG) algorithms; however, both fail to adapt to the complex resource optimization required in the multi-RIS multi-OP network. The low-quality real-time policies at the OP-level directly impact the information matrices uploaded to the RP, further deteriorating RIS allocation and ultimately leading to a catastrophic collapse in overall performance. At the RP-level, we deploy the double deep Q-learning (DDQN) algorithm and attempt to balance the trade-off between exploration and exploitation using $\epsilon_e = \epsilon^e$. However, the DDQN algorithm still struggles to achieve a satisfactory sum-rate. In conclusion, the experimental results further underscore the necessity of the HPPO algorithm.

\subsection{Superiority of HPPO Algorithm}
In this subsection, we conduct a detailed comparison to demonstrate the superior performance of HPPO. To ensure fairness, PPO is deployed in each OP-agent, while the RP-agent implements various RIS allocation benchmarks driven by different information uploaded by the OPs, as listed below.
\begin{enumerate}
    \item \textbf{Auction scheme} \cite{TVT Cai}: At the end of the $e-1$-th episode, each OP calculates the performance gain when each RIS operates independently, and then uploads the private bids to the RP. The RP uses the auction algorithm to determine the RIS allocation for the $e$-th episode (labeled as ``Auc+PPO''). 
    \item \textbf{Distance scheme}: The RP obtains the average distance from each RIS to all users of each OP, and allocates the RIS to the OP with the shortest average distance (labeled as ``NRU+PPO'').
    \item \textbf{Random scheme}: At each episode, the RP does not accept any information uploaded by the OPs and instead randomly allocates RISs to OPs (labeled as ``Rand+PPO'').
    \item \textbf{Greedy scheme}: At the end of the $e-1$-th episode, each OP calculates the sum-rate for each RIS when serving independently and uploads the results to the RP. In the $e$-th episode, the RP allocates the RIS to the OP that can achieve the highest sum-rate (labeled as ``Gre+PPO''). 
    \item \textbf{w/o RIS}: As in previous simulations, there is no RIS (RP) in the environment, and each OP is treated as an independent entity. (labeled as ``PPO'').

\end{enumerate}

\begin{figure}[t]
\centering
\includegraphics[width = 3.0 in]{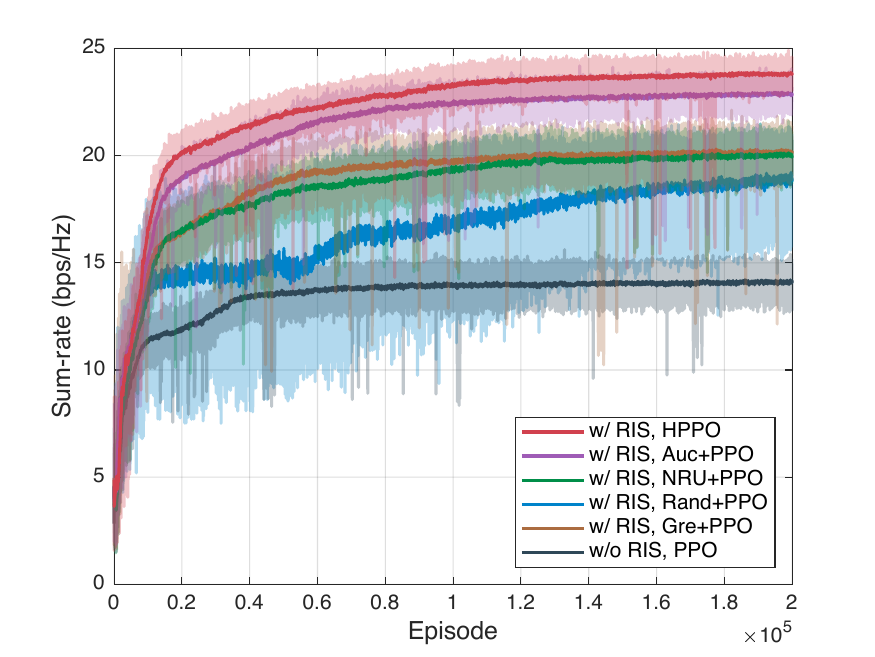}
\caption{Sum-rate convergence for different algorithms.}
\label{fig_curves_benchmarks}
%\vspace{-0.3 cm}
\end{figure}

\begin{figure}[t]
\centering
\includegraphics[width = 3.0 in]{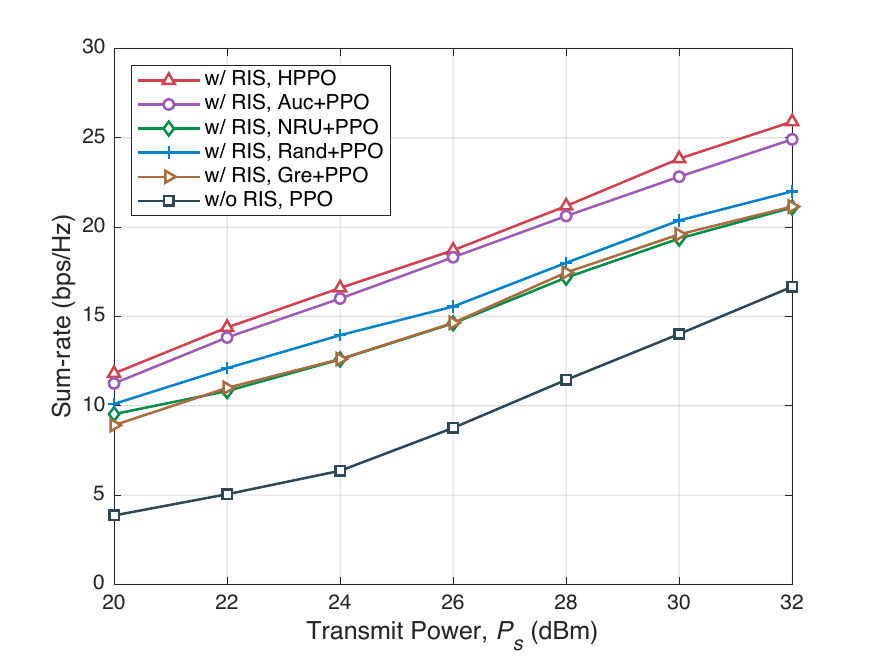}
\caption{Sum-rate versus transmit power.}
\label{fig_rate_vs_P}
%\vspace{-0.3 cm}
\end{figure}

Fig. \ref{fig_curves_benchmarks} clearly illustrates the sum-rate variation curves for different benchmarks, demonstrating that all RIS allocation algorithms can achieve collaborative resource optimization. However, for the auction and greedy schemes, OPs must calculate the performance gain of each RIS individually, requiring additional estimation overhead to obtain RIS-related channels \eqref{eq_subchannel}. For the random scheme, the sum-rate curve shows greater performance fluctuations and takes longer to converge.

\begin{figure}[t]
\centering
\includegraphics[width = 3.0 in]{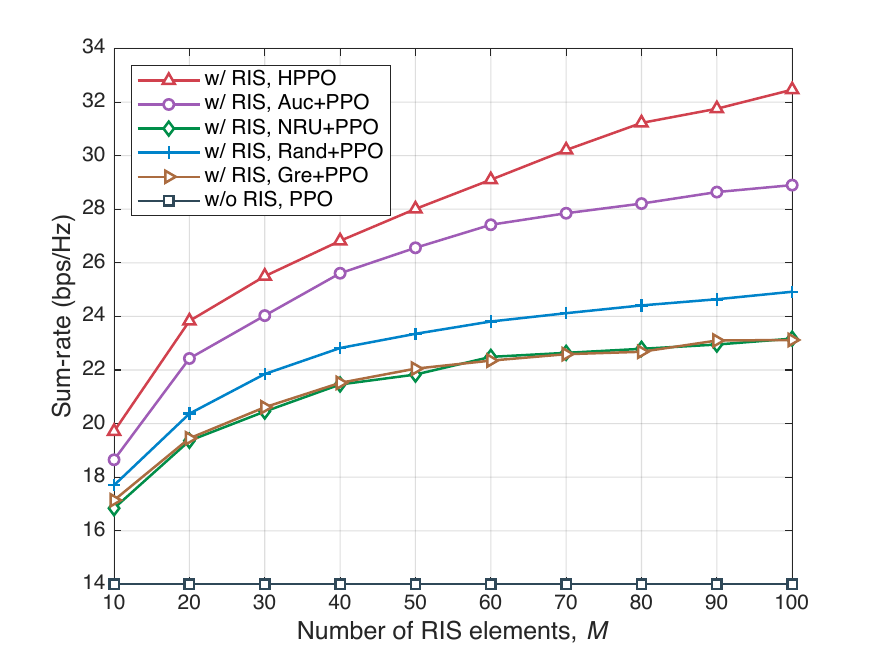}
\caption{Sum-rate versus the number of RIS elements.}
\label{fig_rate_vs_M}
%\vspace{-0.3 cm}
\end{figure}

In Fig. \ref{fig_rate_vs_P}, we show the sum-rate versus transmit powers $P_s$ ($P_{j_s}=P_s / J_s$), highlighting the superior performance of the HPPO algorithm. Specifically, the heuristic auction scheme may lead to locally optimal RIS allocation, thus hindering further resource optimization for the OPs. The distance scheme only uses distance information to output a fixed RIS allocation, lacking effective utilization of environmental information. For the greedy scheme, the RP assigns RISs to the OP that provides the highest sum-rate, but ignores the interference RISs impose on unassociated OPs, significantly impacting overall performance. As for the random scheme, although its performance surpasses some allocation schemes, it suffers from prolonged convergence time and substantial fluctuations, making it unsuitable for practical deployment. In summary, the proposed HPPO algorithm balances RIS collaboration and interference management, consistently achieving the best performance across various HDRL tasks corresponding to different environmental variables.

In Fig. \ref{fig_rate_vs_M}, we show the sum-rate versus number of RIS elements $M_l$. The results indicate that as $M_l$ increases, the sum-rate performance gap between the HPPO algorithm and the other benchmarks gradually widens. This is because the RIS allocation results from the RP determine the upper limit for OP resource optimization. Suboptimal RIS allocations may lead to ineffective utilization of power resources by OPs and increased interference between RISs, further suppressing the optimal performance of all RIS elements. When $M_l=100$, the HPPO algorithm achieved a sum-rate improvement of 14\% compared to the auction scheme, 33\% compared to the random scheme, and nearly 39\% compared to both the greedy and distance schemes.

\begin{figure}[t]
\centering
\includegraphics[width = 3.0 in]{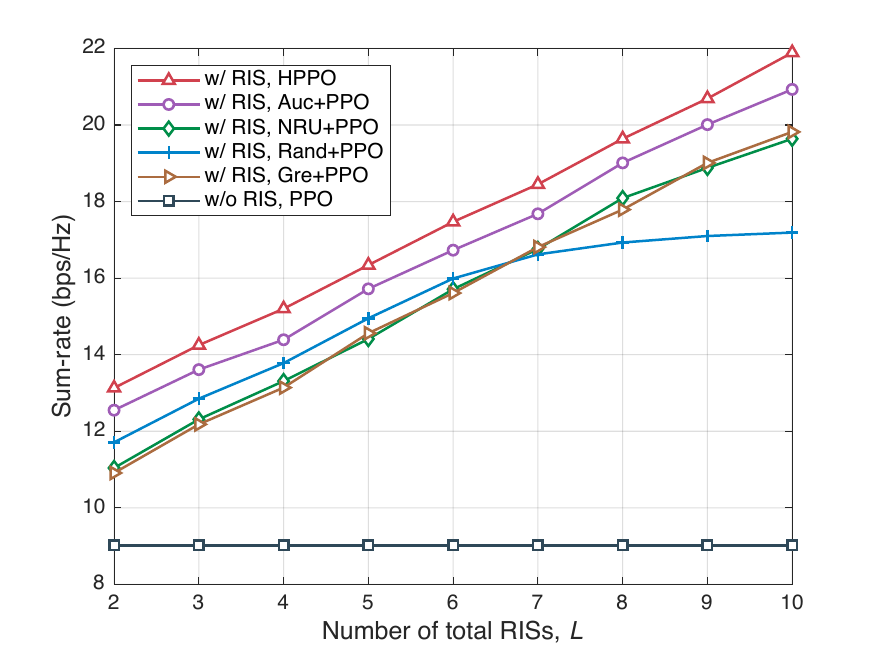}
\caption{Sum-rate versus the number of RISs.}
\label{fig_rate_vs_L}
%\vspace{-0.3 cm}
\end{figure}

In Fig. \ref{fig_rate_vs_L}, we present the sum-rate versus the number of RISs $L$ under $S=2$, $N=4$, and $M_l=10$, where the possible cases for RIS allocation range from $2^2$ to $2^{10}$. Except for the random scheme, which fails to converge effectively and shows performance degradation, all other benchmarks can achieve effective RIS allocation. Specifically, the HPPO algorithm obtains the probability values of all $S^L$ goals and samples the RIS allocation accordingly. Simultaneously, it updates the policy in parallel to balance exploration and exploitation, demonstrating superiority in most real-time multi-RIS multi-OP networks.

\begin{figure}[t]
\centering
\includegraphics[width = 3.0 in]{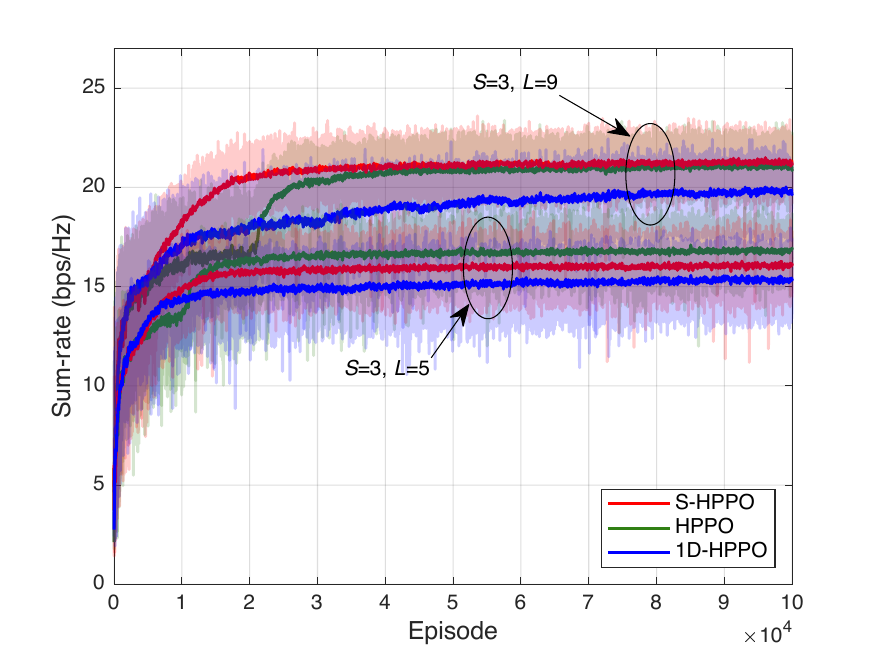}
\caption{Sum-rate convergence for different algorithms.}
\label{fig_curves_SHPPO}
%\vspace{-0.3 cm}
\end{figure}

\begin{figure}[t]
\centering
\includegraphics[width = 3.0 in]{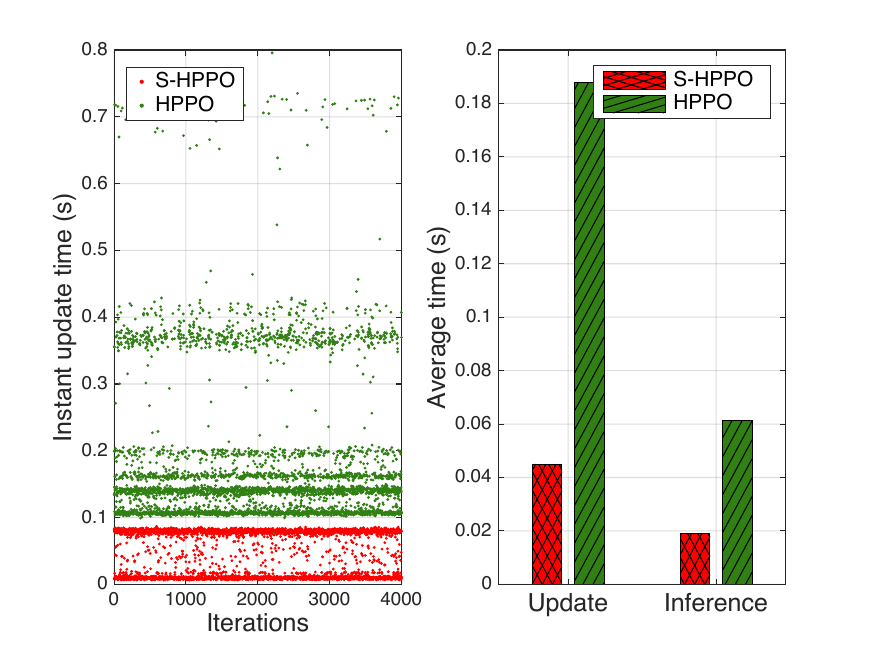}
\footnotesize \hbox{\;\;\;\;(a)  \hspace{2.8cm} (b) }
\caption{Comparison of running time at the RP-level. (a) instant update time. (b) average update and inference time.}
\label{fig_time}
%\vspace{-0.3 cm}
\end{figure}

\begin{figure}[t]
\centering
\includegraphics[width = 3.0 in]{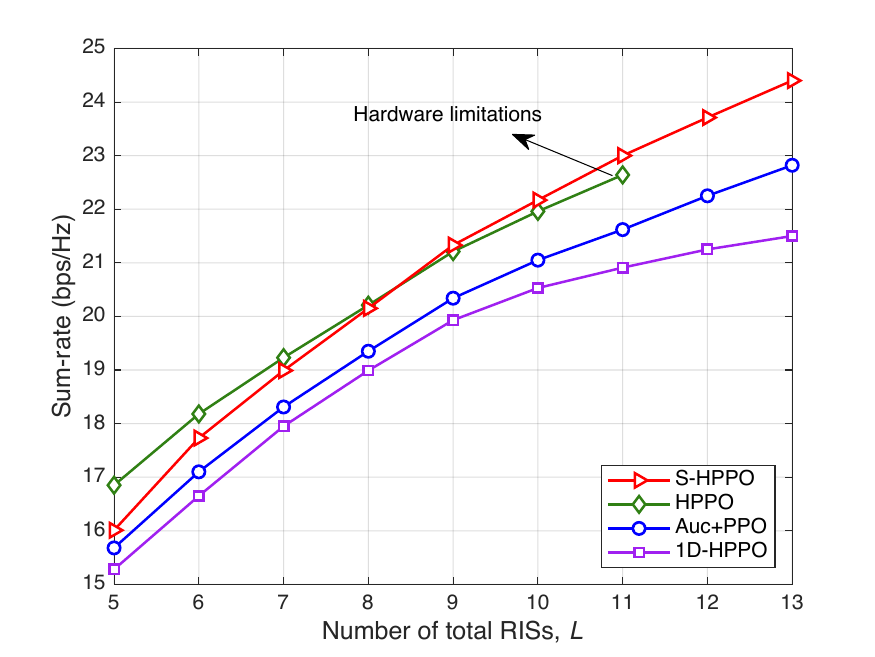}
\caption{Sum-rate versus number of RISs.}
\label{fig_rate_SHPPO}
%\vspace{-0.3 cm}
\end{figure}

\subsection{Effectiveness of S-HPPO Algorithm for Large-Scale RIS Allocation}

In this subsection, we provide a detailed comparison between the S-HPPO and HPPO algorithms to further analyze the impact of the exponentially expanding RIS allocation space on overall sum-rate performance. For comparison, we employ a common approach for handling high-dimensional discrete spaces, where the RP-agent outputs a continuous value in the range of (-1, 1), which is then quantized into the discrete space of $S^L$ (labeled as ``1D-HPPO''). Simulations are conducted under $M_l=10$, $K_s=2$, and $N=4$.

In Fig. \ref{fig_curves_SHPPO}, we present the sum-rate variation for three algorithms under $S=3, L=5$ and $S=3, L=9$. The results show that the 1D-HPPO approach does not achieve better convergence speed or sum-rate performance. This is due to the extremely fine quantization steps, which result in unstable RIS allocation outputs and increase the likelihood of the algorithm getting trapped in local optima. Furthermore, in Fig. \ref{fig_time}, we compare the runtime of S-HPPO and HPPO under $S = 5, L = 9$. Specifically, Fig. \ref{fig_time} (a) compares the update times for 4000 iterations, while Fig. \ref{fig_time} (b) compares the average update and inference times. The experimental results clearly demonstrate the superior update efficiency of S-HPPO in large-scale RIS allocation scenarios, with a 318\% improvement in updating and a 221\% improvement in inferencing. In summary, the S-HPPO algorithm ensures a relatively stable convergence speed across RIS allocation tasks of varying scales. In contrast, as $L$ increases, the centralized output goal space of HPPO at the RP-level significantly slows down both convergence and update speed.

In Fig. \ref{fig_rate_SHPPO}, we present the  sum-rate performance of various algorithms with $S=3$ as $L$ increases from 5 to 13, corresponding to an output goal space ranging from $3^5$ to $3^{13}$. Experimental results show that when the output goal space is smaller than $3^9$, the performance of HPPO still surpasses that of S-HPPO. This is because the approximate decomposition of the RP-level MDP in S-HPPO leads to a slight performance degradation. 
However, as $L$ increases, the dimensionality issue of the goal space becomes more severe, narrowing the performance gap between S-HPPO and HPPO, with S-HPPO surpassing HPPO when $L > 9$.  Additionally, when $L=11$, the number of parameters at the RP-level in HPPO reaches the magnitude of $10^8$, hitting the hardware limit for real-time policy optimization. Therefore, for large-scale RIS allocation, S-HPPO emerges as a superior choice owing to its superior performance, faster convergence, and reduced hardware constraints compared to HPPO. Furthermore, the performance gap between the auction strategy and 1D-HPPO and S-HPPO continues to widen, highlighting that both traditional quantization scheme and auction scheme are increasingly affected by the dimensionality issue.

\section{Conclusions}\label{section:7}

In this paper, we proposed an HDRL approach for resource optimization in multi-RIS multi-OP networks. Within the proposed HDRL framework, the RP designed the current episode's RIS allocation based on the historical information matrix uploaded by each OP, while the OPs utilized historical equivalent CSI to determine user association, beamforming, and the phase-shifts of the assigned RISs. Simulation results demonstrated that the HPPO algorithm could achieve joint resource optimization for both the RP and OPs across various environments, while also exhibiting superior performance. In scenarios with 2 OPs, the presence of the RP yielded sum-rate gains of 85\% and 57\% for each OP. Additionally, the improved S-HPPO algorithm was better suited for large-scale RIS allocation environments. In the future, we will explore more dynamic mobile user scenarios and take the diverse needs of OPs into account within the RP.

\end{document}